\documentclass[11pt]{article}

\usepackage[T1]{fontenc}
\usepackage[utf8]{inputenc}
\usepackage[margin=1in]{geometry}

\usepackage{authblk}

\newcommand{\equalcontrib}{\textsuperscript{\dag}}
\newcommand{\corrauthor}[1]{\thanks{Corresponding author: #1}}

\usepackage{graphicx}
\usepackage{dcolumn}
\usepackage{bm}
\usepackage{amsmath,amssymb}
\usepackage{xcolor}
\usepackage{soul}
\usepackage{nicefrac}
\usepackage{caption}
\usepackage{comment}
\usepackage{siunitx}
\usepackage[numbers,sort&compress]{natbib}
\usepackage[hidelinks]{hyperref}
\usepackage{xr}

\newcommand{\didv}{{d}\textit{I}/{d}\textit{V}}
\newcommand{\LI}{$V_{\text{m}}$}

\makeatletter
\newcommand*{\addFileDependency}[1]{% argument=file name and extension
  \typeout{(#1)}
  \@addtofilelist{#1}
  \IfFileExists{#1}{}{\typeout{No file #1.}}
}
\makeatother
\newcommand*{\myexternaldocument}[1]{%
    \externaldocument{#1}%
    \addFileDependency{#1.tex}%
    \addFileDependency{#1.aux}%
}
\myexternaldocument{Supplemental}

\title{Engineering Ferrimagnetic Interactions in Molecular Quantum Systems}

\title{Engineering Ferrimagnetic Interactions in Molecular Quantum Systems}

\author[1,7]{Elia Turco\equalcontrib\corrauthor{e.turco@tudelft.nl}}
\author[2]{Fupeng Wu\equalcontrib}
\author[3]{Annika Bernhardt\equalcontrib}
\author[1]{Nils Krane}
\author[4,5]{Ji Ma}
\author[1,6]{Roman Fasel}
\author[3]{Michal Jur\'{\i}\v{c}ek\corrauthor{michal.juricek@chem.uzh.ch}}
\author[2]{Xinliang Feng\corrauthor{xinliang.feng@tu-dresden.de}}
\author[1]{Pascal Ruffieux\corrauthor{pascal.ruffieux@empa.ch}}

\affil[1]{nanotech@surfaces Laboratory, Empa---Swiss Federal Laboratories for Materials Science and Technology, 8600 D\"{u}bendorf, Switzerland}
\affil[2]{Max Planck Institute of Microstructure Physics, Weinberg 2, 06120 Halle, Germany; and Center for Advancing Electronics Dresden (cfaed) \& Faculty of Chemistry and Food Chemistry, Technische Universit\"at Dresden, 01062 Dresden, Germany}
\affil[3]{Department of Chemistry, University of Zurich, 8057 Zurich, Switzerland}
\affil[4]{College of Materials Science and Optoelectronic Technology \& Center of Materials Science and Optoelectronics Engineering, University of Chinese Academy of Sciences, 100049 Beijing, P. R. China}
\affil[5]{Beijing National Laboratory for Molecular Science, CAS Key Laboratory for Organic Solids, Institute of Chemistry, Chinese Academy of Sciences, Beijing 100190, China}
\affil[6]{Department of Chemistry, Biochemistry and Pharmaceutical Sciences, University of Bern, 3012 Bern, Switzerland}
\affil[7]{Current address: QuTech and Kavli Institute of Nanoscience, Delft University of Technology, 2600 GA Delft, The Netherlands}

\date{}

\begin{document}
\maketitle

\begin{center}
{\small\textsuperscript{\dag}These authors contributed equally to this work.}
\end{center}

\vspace{1em}

%%%%%%% ABSTRACT %%%%%%%
\begin{abstract}
Achieving long-range ferrimagnetic order in purely organic systems remains a major challenge in molecular magnetism. Here we report the synthesis and characterization of heterospin-coupling motifs, formed by covalently linking spin-\nicefrac{1}{2} and spin-1 triangular nanographenes. A combined solution-phase and on-surface synthetic strategy yields three distinct compounds, whose structures are elucidated by bond-resolved scanning probe microscopy. Starting from a spin-\nicefrac{1}{2}--spin-1 dimer as the elemental ferrimagnetic unit, we employ inelastic electron tunneling spectroscopy to resolve low-energy magnetic excitations and extract the parameters of the Heisenberg Hamiltonian. Extension to trimeric architectures results in two distinct spin configurations, with compensated ($S=0$) and uncompensated ($S=\nicefrac{3}{2}$) ferrimagnetic ground states. The Heisenberg model accurately describes all magnetic transitions, offering direct insight into increasingly complex spin Hamiltonians. These findings establish a molecular platform for designing tunable heterospin systems with robust exchange interactions, opening routes toward multi-level spin encoding in qudit-based quantum technologies.
\end{abstract}

\newpage
%%%%%%% MAIN TEXT %%%%%%%
\section*{Introduction}
\label{introduction}

Organic magnetic materials are an emerging platform for spintronics and quantum computing, enabling quantum-state control at the molecular scale.\cite{gatteschi_magnetic_1991,gaita-arino_molecular_2019,bogani_molecular_2008,sanvito_molecular_2011} Their tunable structure-quantum properties,\cite{zeng_open-shell_2021} longer spin coherence times,\cite{lombardi_synthetic_2021} and potential for sustainable production\cite{pokhodnya_thin-film_2000} make them attractive alternatives to inorganic magnetic systems. However, their practical deployment is hindered by a prevailing tendency toward antiferromagnetic spin coupling, often resulting in complete spin compensation and zero net magnetization. Ferrimagnetic coupling---where spins of unequal multiplicities align antiferromagnetically to yield a net magnetic moment\cite{neel_antiferromagnetism_1952}---offers a compelling alternative, combining the fast spin dynamics and transport characteristics of antiferromagnets  with magnetic field addressability akin to ferromagnets.\cite{kim_ferrimagnetic_2022,finley_spintronics_2020,zhang_ferrimagnets_2023}

 \begin{figure}
\begin{center}
\includegraphics[width=8.6cm]{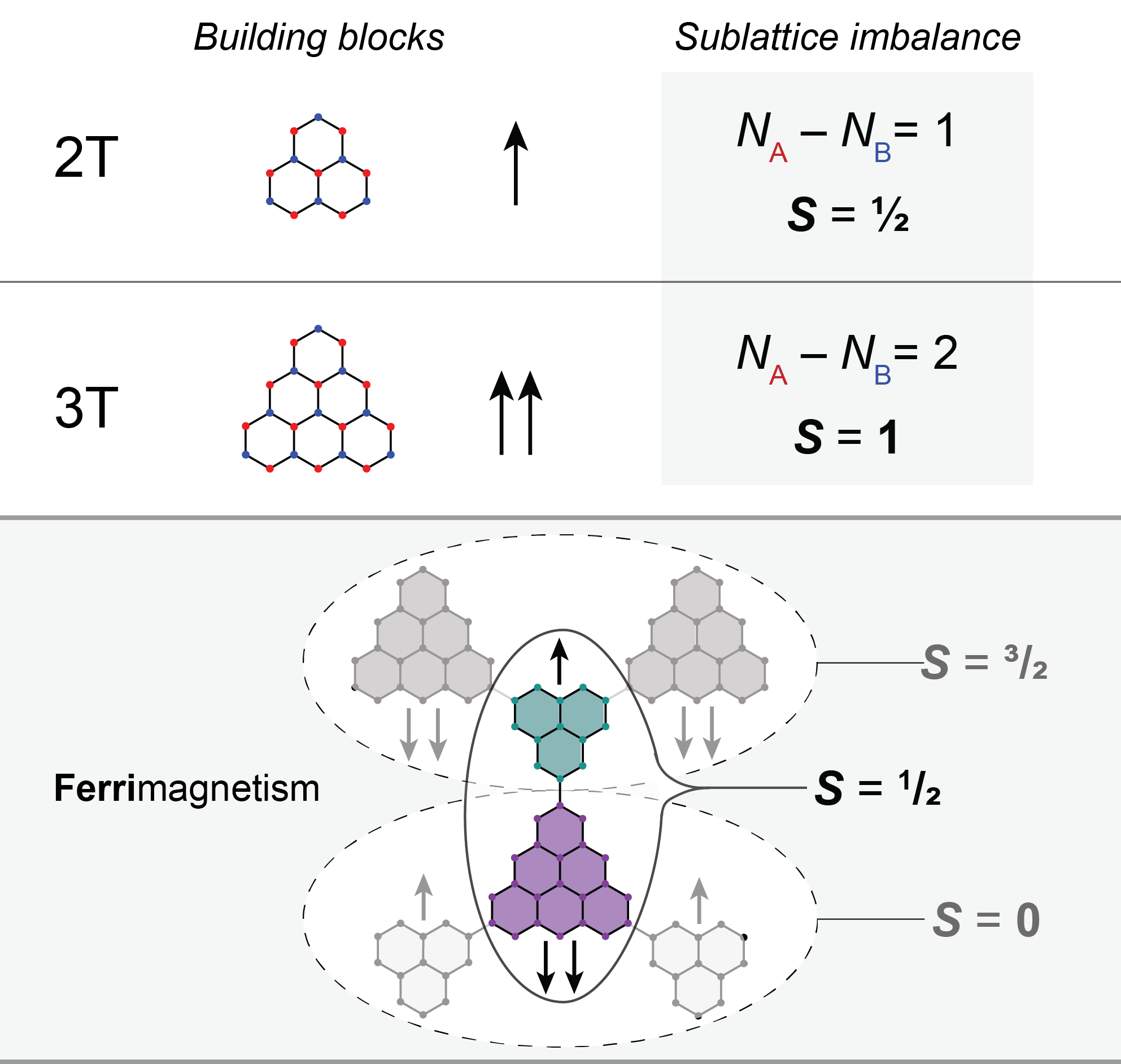}
\caption{Phenalenyl (2T) and [3]triangulene (3T) as magnetic building blocks with spin quantum numbers $S = \nicefrac{1}{2}$ and $S = 1$, respectively. The total spin $S$ arises from the sublattice imbalance, as described by the Ovchinnikov rule\cite{ovchinnikov_multiplicity_1978,lieb_two_1989}, $S = (N_A - N_B)/2$, where $N_A$ and $N_B$ are the number of atomic sites in sublattices A (red) and B (blue). Covalent heterospin coupling of 2T and 3T units into dimers and trimers enables the realization of distinct ferrimagnetic ground states. 
}
\label{Scheme1}
\end{center}
\end{figure}

\begin{figure}
\begin{center}
\includegraphics[width=17.4cm]{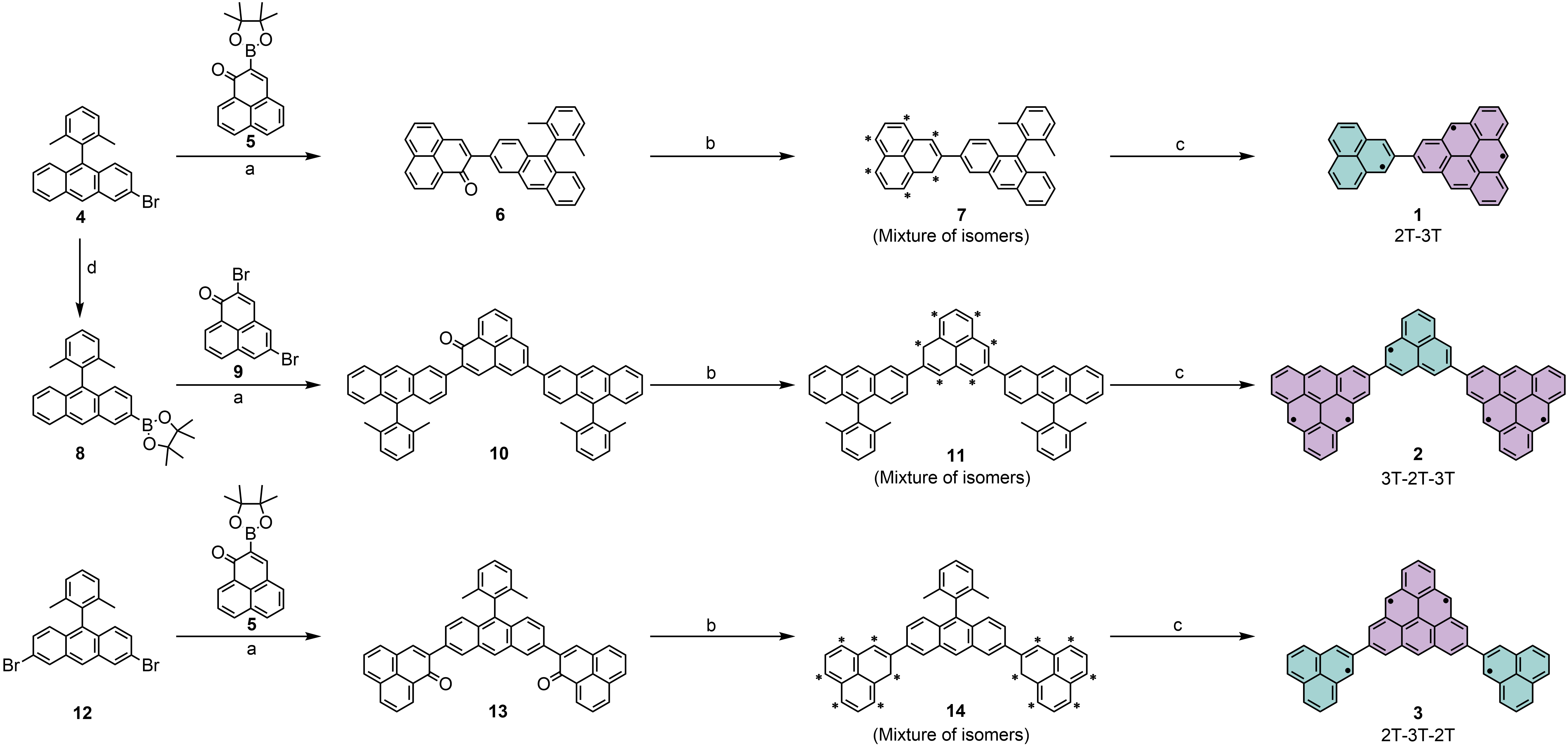}
\caption{Synthetic route to phenalenyl-triangulene dimer and trimers (compounds \textbf{1}-\textbf{3}). Reagents and conditions: (a) Pd(dppf)Cl\textsubscript{2}·CH\textsubscript{2}Cl\textsubscript{2}, K\textsubscript{3}PO\textsubscript{4}, 1,4-dioxane, 85~\textdegree C, 16 h; yields: 90\% and 89\% for compounds \textbf{6} and \textbf{13}, respectively; Pd(PPh\textsubscript{3})\textsubscript{4}, K\textsubscript{2}CO\textsubscript{3}, 1,4-dioxane/H\textsubscript{2}O, 85~\textdegree C, overnight, for compound \textbf{10}, in 79\% yield. (b) DIBAL-H, toluene, rt or 100~\textdegree C, overnight, yields: 32\%, 68\%, and 72\% for compounds \textbf{7}, \textbf{11}, and \textbf{14}, respectively. (c) on Au (111), 320~\textdegree C. (d) 4,4,4',4',5,5,5',5'-octamethyl-2,2'-bi(1,3,2-dioxaborolane), Pd(OAc)\textsubscript{2}, SPhos, K\textsubscript{3}PO\textsubscript{4}, 1,4-dioxane, 90~\textdegree C, overnight, 92\%. 
\textit{Note: In a mixture of isomers, the methylene groups can be shifted to any $\alpha$-position (asterisks) of the phenalenyl subunit.}}
\label{scheme: synthesis}
\end{center}
\end{figure} 

 Since Buchachenko’s seminal proposal in 1979,\cite{buchachenko_unknown_1979} the quest for purely organic ferrimagnets has remained a central challenge in molecular magnetism.
 While long-range ferrimagnetic order is well established in metal–organic systems,\cite{bulled_percolation-induced_2024,manriquez_room-temperature_1991,caneschi_structure_1988,fegy_1d_1998} fully organic analogues remain elusive, limited by weak exchange interactions, low ordering temperatures, and stability problems.\cite{hayakawa_stable_2007,shiomi_single-component_2001,hosokoshi_approach_2001,kanaya_single-component_2001,kaneda_stable_2003,izuoka_magnetically_1994}

Graphene-based $\pi$-electron magnets represent a promising route to overcome these limitations, offering strong and tunable exchange couplings (up to hundreds of meV),\cite{mishra_topological_2020,turco_-surface_2021,barragan_strong_2025,mishra_large_2021, turco_magnetic_2024} along with precise control over spin states and molecular architecture. On-surface synthesis of open-shell nanographenes on coinage metals has enabled the construction of molecular quantum spin chains, including homospin $S = 1$ and $S = \nicefrac{1}{2}$ architectures,\cite{mishra_observation_2021,sun_-surface_2025,zhao_spin_2025} offering an ideal platform to explore quantum magnetism and topological phases in low-dimensional systems.

 Here, we take a first step toward extending this strategy to heterospin  systems\cite{yamaguchi_experimental_2020,ahami_quantum_2024} by synthesizing three model compounds featuring an alternating spin-$1$/spin-$\nicefrac{1}{2}$ motif.

In contrast to the extensively studied homospin configurations,\cite{mishra_collective_2020,krane_exchange_2023,turco_magnetic_2024,hieulle_-surface_2021,song_highly_2024,perez-elvira_reactivity_2025} heterospin coupling remains largely unexplored, with previous realizations limited to atomically precise metal atom assemblies\cite{otte_spin_2009,muenks_correlation-driven_2017,yang_coherent_2019} or metal–organic hybrids on surfaces.\cite{huang_quantum_2025,zhang_atomic-scale_2025}

To realize a heterospin coupling motif in a purely organic framework, we combine two nanographene building blocks with distinct spin multiplicities. This strategy exploits the bipartite structure of the graphene honeycomb lattice, where sublattice imbalance dictates the total spin quantum number as $S = (N_A - N_B)/2$.\cite{ovchinnikov_multiplicity_1978,lieb_two_1989}

%\noindent
The two smallest members of the triangulene family---namely phenalenyl (2T, $S=\nicefrac{1}{2}$) and [3]triangulene (3T, $S=1$)---are herein employed as magnetic building blocks (Scheme \ref{Scheme1}). Covalent bonding at the $\beta$-positions (minority sublattices) of 2T and 3T yields strong antiferromagnetic exchange, mediated by third-nearest-neighbor hopping ($t_3$).\cite{jacob_theory_2022,krane_exchange_2023} The 2T–3T dimer (\textbf{1}, Scheme \ref{scheme: synthesis}) represents the fundamental ferrimagnetic unit, synthesized via a combined solution-phase and on-surface synthesis on Au(111). Using the same synthetic strategy, we obtained two trimeric compounds: 3T–2T–3T (\textbf{2}) and 2T–3T–2T (\textbf{3}), featuring quartet and singlet ground states, respectively. A combination of low-temperature (4.5 K) scanning tunneling microscopy (STM), atomic force microscopy (AFM), and scanning tunneling spectroscopy (STS) reveals spin excitations in \textbf{1}–\textbf{3} that are accurately captured by a minimal Heisenberg model.\cite{ternes_probing_2017} These findings establish a modular platform for engineering all-carbon ferrimagnets and underscore their potential as molecular qudits in quantum information applications.\cite{atzori_second_2019,chiesa_theoretical_2022,wernsdorfer_synthetic_2019,gaita-arino_molecular_2019}

%We demonstrate that covalent coupling of molecular units with distinct spin multiplicities yields molecular quantum systems with more than two well-defined spin states ($d > 2$), commonly referred to as qudits in the context of quantum information. The resulting expansion of the accessible Hilbert space enables information encoding beyond binary logic, offering a compact and scalable platform for advanced quantum operations\cite{atzori_second_2019,chiesa_theoretical_2022,wernsdorfer_synthetic_2019,gaita-arino_molecular_2019}.

\section*{Results and Discussion}
\label{results_discussion}

\begin{figure*}[h!]
\begin{center}
\includegraphics[width=17.4cm]{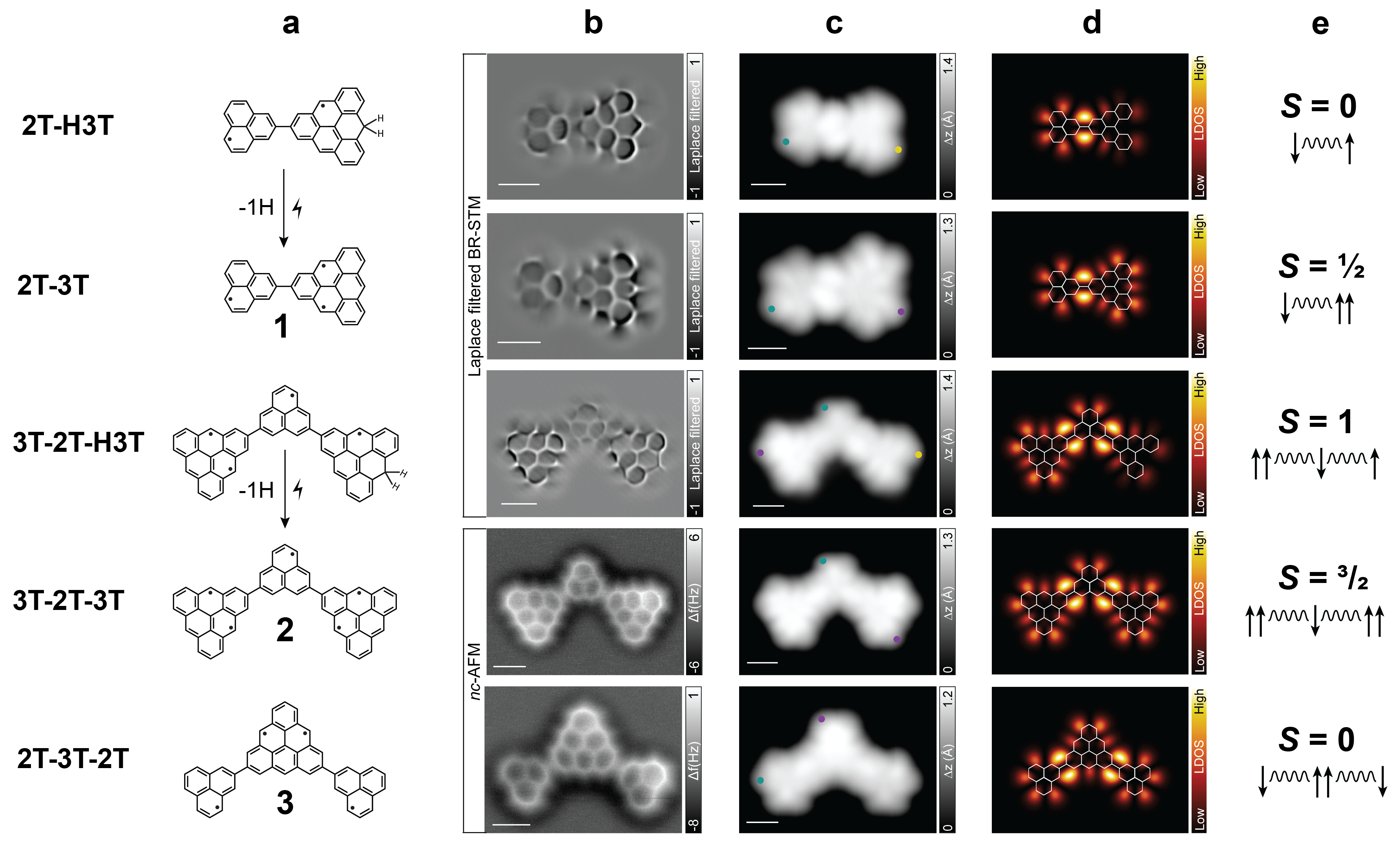}
\caption{On-surface synthesis of dimers and trimers composed of covalently coupled 2T and 3T building blocks. (a) Chemical sketch of the nanographenes synthesized and characterized in this work. Molecules labeled as \textbf{1}-\textbf{3} correspond to the target compounds obtained via thermal annealing of their respective \textbf{7}, \textbf{11}, and \textbf{14} precursors, while 2T-H3T and 3T-2T-H3T are are identified as hydro intermediates. (b) Structural characterization of molecules in (a), carried out with a carbon monoxide (CO) functionalized tip by using bond-resolved STM (opened feedback on the molecule with -5 mV/50 pA, $\Delta z=-0.7 $\AA ) or \textit{nc}-AFM techniques (open-feedback on Au(111) with -5 mV/100 pA, $\Delta z = 1.9-2.0$ \AA). (c) High-resolution STM images, closed feedback  with -0.1 V/100-150 pA, acquired with a CO-functionalized tip. Filled colored circels indicate the positions where the spectra in Figures \ref{fig: magnetic dimers} and \ref{fig:magnetic trimers} were acquired. (d) Calculated MFH-LDOS maps of the spin-carrying orbitals for each molecular structure, with molecular skeletons overlaid. The CH$_2$ group is incorporated into the MFH model by removing the corresponding carbon site from the $\pi$-system. (e) Total spin quantum number $S$ and the corresponding Heisenberg spin model for structures in (a). Scale bars: 0.5 nm (b,c).     }
\label{structural characterization}
\end{center}
\end{figure*}

The 2T--3T dimer (\textbf{1}), 3T--2T--3T trimer (\textbf{2}), and 2T--3T--2T trimer (\textbf{3}) were synthesized via surface-assisted cyclodehydrogenation of molecular precursors \textbf{7}, \textbf{11}, and \textbf{14}. These precursors were obtained in solution through a two-step synthetic route, as illustrated in Scheme~\ref{scheme: synthesis} (detailed procedures and characterization data are provided in the Supporting Information). In the first step, a Suzuki coupling reaction between 2-bromo-10-(2,6-dimethylphenyl)anthracene (\textbf{4}) and 2-(4,4,5,5-tetramethyl-1,3,2-dioxaborolan-2-yl)-1\textit{H}-phenalen-1-one (\textbf{5}) afforded 2-(10-(2,6-dimethylphenyl)anthracen-2-yl)-1\textit{H}-phenalen-1-one (\textbf{6}) in 90\% yield. Subsequent reduction of \textbf{6} with diisobutylaluminum hydride (DIBAL-H) yielded the 2T-3T precursor \textbf{7}
as a mixture of isomers where the methylene groups can be shifted to any $\alpha$-position (asterisks) of the phenalenyl subunit. 
%Due to the instability of the individual isomers and the difficulty of their isolation, this mixture was directly employed as the precursor for on-surface synthesis of dimer \textbf{1} on Au(111) surface via annealing at 320\,\textdegree{}C. 
Following a similar strategy, Suzuki coupling between 2-(10-(2,6-dimethylphenyl)anthracen-2-yl)-4,4,5,5-tetramethyl-1,3,2-dioxaborolane (\textbf{8}) and 2,5-dibromo-1\textit{H}-phenalen-1-one (\textbf{9}) afforded 2,5-bis(10-(2,6-dimethylphenyl)anthracen-2-yl)-1\textit{H}-phenalen-1-one (\textbf{10}) in 79\% yield. Reduction of \textbf{10} with DIBAL-H gave an isomeric mixture (\textbf{11}), which was used as the precursor for trimer \textbf{2}. Similarly, Suzuki coupling of compound \textbf{12} with compound \textbf{5} afforded 2,2'-(10-(2,6-dimethylphenyl)anthracene-2,7-diyl)bis(1\textit{H}-phenalen-1-one) (\textbf{13}) in 89\% yield. Reduction of \textbf{13} with DIBAL-H afforded an isomeric mixture (\textbf{14}), which served as the precursor for trimer \textbf{3}. 

The on-surface synthesis of \textbf{1}, \textbf{2} and \textbf{3} was achieved in two steps.  First, the precursors \textbf{7}, \textbf{11}, and \textbf{14} were independently deposited onto atomically clean Au(111) surfaces at room temperature; STM analysis of the resulting adsorbed molecular species is provided in Figure~S2. Subsequent annealing to 320~\textdegree C triggered oxidative ring closure of the methyl groups, yielding the 2T--3T dimer \textbf{1} and the two trimers 3T--2T--3T (\textbf{2}) and 2T--3T--2T (\textbf{3}), as shown in the overview STM images in Figure~S3.

Single-molecule STM and bond-resolved STM/AFM imaging (Figure~\ref{structural characterization}c,b) confirm the successful on-surface synthesis of target structures \textbf{1}–\textbf{3}, along with hydro intermediates 2T-H3T and 3T-2T-H3T. The latter likely result from partial passivation of radical centers via substitution of CH groups with CH\textsubscript{2} groups, attributed to atomic hydrogen diffusion during annealing. The corresponding chemical structures (Figure~\ref{structural characterization}a) define five distinct spin Hamiltonians, with total spin quantum numbers ranging from $S=0$ to $S=\nicefrac{3}{2}$, modeled using the Heisenberg formalism (Figure~\ref{structural characterization}e). To validate the magnetic ground state assignments, we employed a tight-binding framework with electron correlation treated at the mean-field Hubbard level (MFH-TB), and computed the local density of states (LDOS) for each structure. 
\begin{figure}[h!]
    \centering
    \includegraphics[width=8.6cm]{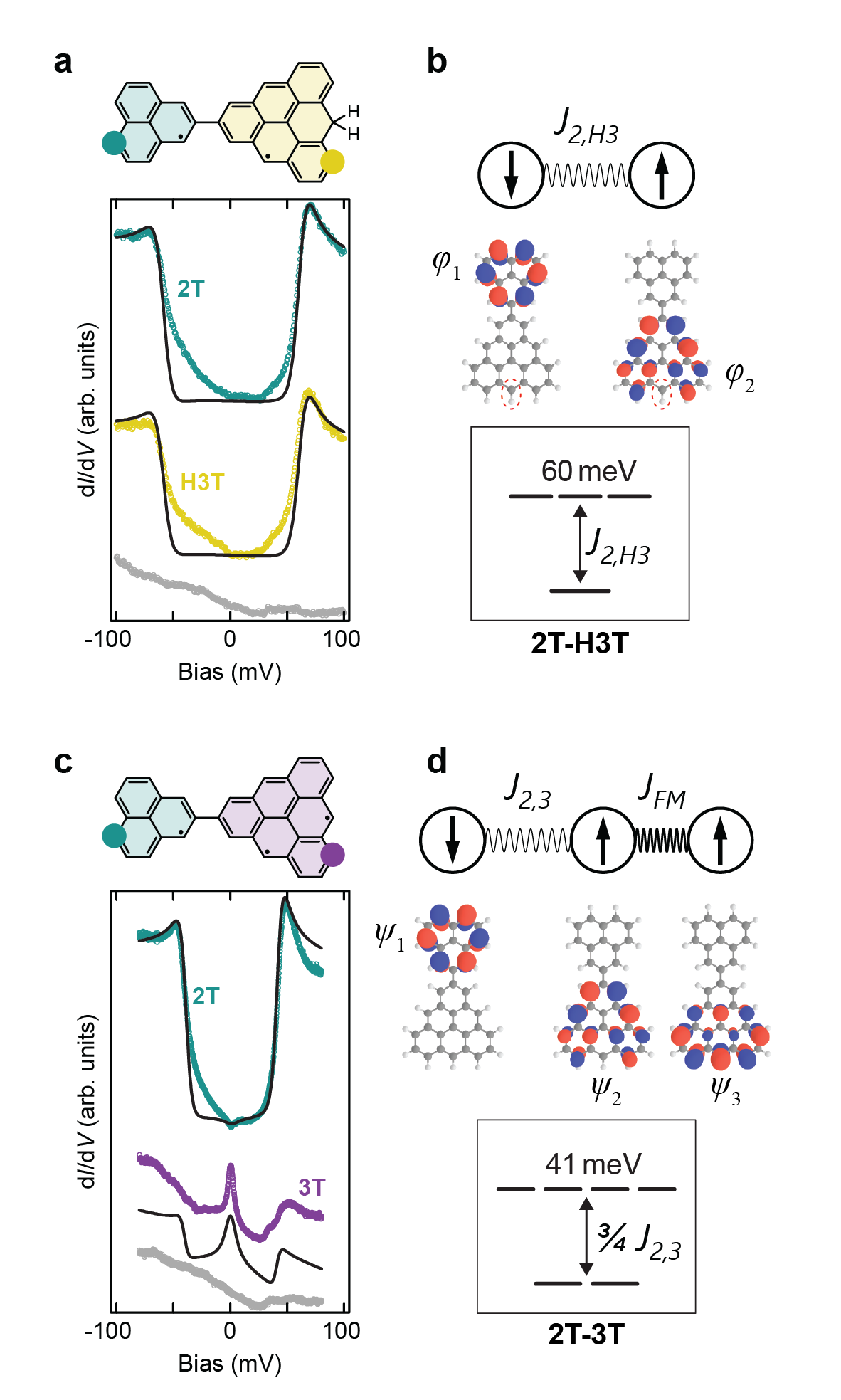}
    \caption{Extraction of spin Hamiltonian parameters from STS spectra.
    (a) Experimental and simulated differential conductance \didv spectra of 2T--H3T.
    The spectra were taken at positions marked on the sketch (top) and in Figure~\ref{structural characterization}c. The feedback loop was opened on the molecule at -0.1~V/850~pA (a) and -80~mV/850~pA (b).  Lock-in modulation voltage \LI~=~1~mV (a,b).     
    The simulated spectra were calculated by a third order scattering model assuming Heisenberg coupling, as depicted in (b) and described in the Supporting Information.
    The frontier states corresponding to the spins in the Heisenberg model are shown as iso-surfaces below.
    Matching the simulated spectra with the experiment yields an effective coupling of $J_{2,H3}=60\,\text{meV}$.
    (c-d) same as (a-b) but for 2T--3T. Assuming the ferromagnetic coupling of the two 3T spins to be much larger than the coupling to 2T ($J_\text{FM} \gg J_{2,3}$) the observed doublet--quartet gap corresponds to $\frac{3}{4}J_{2,3} \approx 41\,\text{meV}$, thus $J_{2,3} = 54\,\text{meV}$.
}
    \label{fig: magnetic dimers}
\end{figure}

\begin{figure}[h!]
    \centering
    \includegraphics[width=8.6cm]{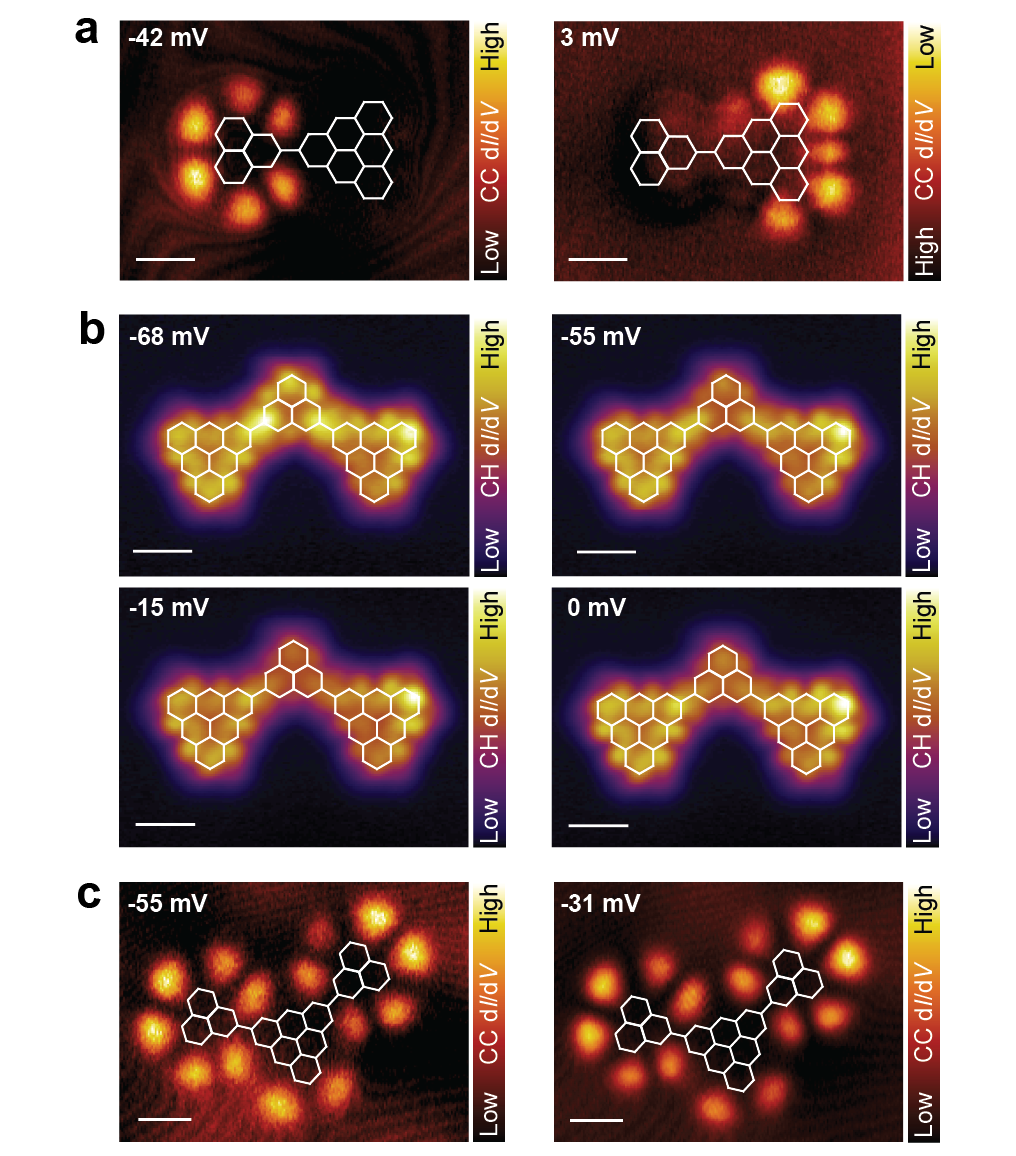}
    \caption{Spin excitation maps of \textbf{1}-\textbf{3}. (a,c) Constant-current (CC) maps and (b) constant-height (CH) maps acquired at the voltage thresholds corresponding to the spin excitation energies (maps at positive biases are provided in the Supporting Information). CC setpoints: (a)-left 1 nA, (a)-right 100 pA, (c) 350 pA. CH maps (b): feedback opened on the molecule at -0.1 V/800 pA. For the map on the right of panel (a), depicting the spatial extension of the Kondo-like resonance, we have leveraged the topographic effect of constant-current maps. At small finite bias, the presence of a zero-bias peak leads to tip retraction, thereby reducing the measured \didv~intensity. To enhance the visibility of the Kondo resonance, the color scale has been inverted (original map is reported in the Supporting Information). Lock-in modulation \LI: (a, c) 2 mV , (b,d) 4 mV. (a,c) were acquired with a metal tip, (b) with a CO-functionalized tip. Scale bars: 0.5 nm.    }
    \label{fig:spin excitations}
\end{figure}
 The resulting LDOS maps of spin-carrying orbitals (Figure~\ref{structural characterization}d) show excellent agreement with the apparent topography of the in-gap STM images acquired at $V=-0.1$ V (Figure~\ref{structural characterization}c), reflecting the spatial distribution of the orbitals involved in low-energy spin excitations.
While the following section focuses on the magnetic properties, the electronic structure of \textbf{1}, \textbf{2}, and \textbf{3} was also investigated by detailed STS and MFH-TB analysis (Figures~S4, S7 and S8).
The observed features highlight the many-body character of the coupled open-shell nanographenes (Figure~S5), in line with previous studies on related systems.\cite{krane_exchange_2023}

\subsection*{Magnetic characterization}

\begin{figure*}
    \centering
    \includegraphics[width =17.4cm]{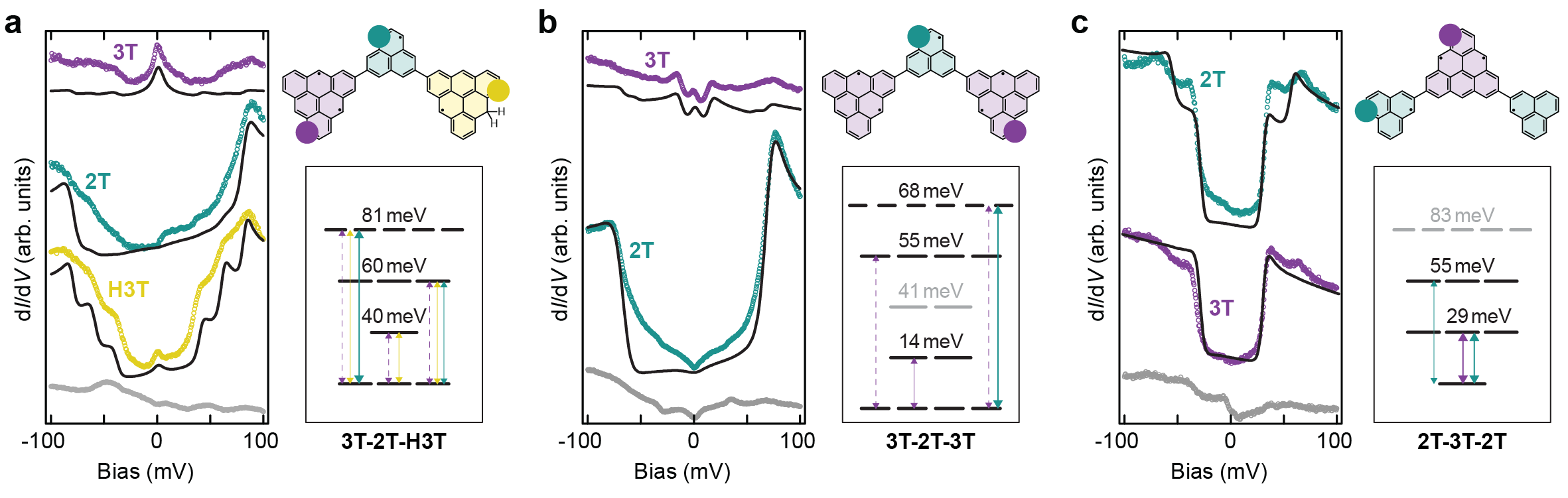}
    \caption{Experimental and calculated IETS spectra for \textbf{2} and \textbf{3}.
    (a) Differential conductance spectra taken at three different positions on 3T--2T--H3T, marked as on the molecule sketch and in Figure~\ref{structural characterization}c. The feedback loop was opened on the molecule at -0.1~V/800~pA (a,b) and -0.1/500~pA (c).  Lock-in modulation voltage \LI~=~2~mV (a,b) and \LI~=~1~mV (c).
    Black solid lines correspond to simulated spectra using the coupling parameters $J_{2,3}$ and $J_{2,H3}$ as determined before.
    The energy scheme shows the available spin excitated states, whereas the multiplicity is represented by the number of lines.
    Colored arrows indicate the excitation intensity, depending on tip position. %(b) and (c) the same as (a) for 3T--2T--3T and 2T--3T--2T, respectively.
    }
    \label{fig:magnetic trimers}
\end{figure*}

We now turn to the low-bias STS analysis of the structures presented in Figure~\ref{structural characterization}a, with the aim of validating the previously assigned Heisenberg representation (Figure~\ref{structural characterization}e). To this end, we first determine the spin Hamiltonian parameters from the dimeric coupling motifs (Figure~\ref{fig: magnetic dimers}), and then use the as-determined values to model the spin excitations observed in the trimeric systems (Figure~\ref{fig:magnetic trimers}).

Antiferromagnetic coupling of two spin-\nicefrac{1}{2} units is realized via the structure 2T-H3T, where the additional hydrogen atom effectively removes one unpaired electron from the triangulene unit, with the remaining unpaired $p_z$ electron delocalized over the triangulene backbone. The spatial distribution of the resulting spin-carrying orbitals, $\varphi_1$ and $\varphi_2$, obtained from TB-MFH calculations, is shown in Figure~\ref{fig: magnetic dimers}b. Low-bias STS spectra acquired on the 2T and H3T units (Figure~\ref{fig: magnetic dimers}a) reveal two symmetric steps around the Fermi level, corresponding to inelastic singlet--triplet excitations. These features are well reproduced by a Heisenberg dimer model, with Hamiltonian $\mathcal{H}=J_{2,\mathrm{H3}}\, \mathbf{S}_1 \cdot \mathbf{S}_2$, including spin-flip processes up to third order.\cite{ternes_spin_2015} 

The corresponding calculated spectra, shown in black in the graph, closely reproduce the experimental features with an effective exchange coupling of $J_{2,\mathrm{H3}} =60$ meV---significantly larger than values reported for symmetric dimers.\cite{krane_exchange_2023,catarina_conformational_2024}. To rationalize this, we estimate the effective hopping ($t_{\mathrm{eff}}$) and Coulomb repulsion ($U_{\mathrm{eff}}$), which enter the expression for the exchange coupling as $J = 4t_{eff}^2/U_{eff}$, using the TB-MFH model.\cite{jacob_theory_2022} 
Although $t_{\mathrm{eff}}$ is only $5\%$ larger than in the 2T–2T dimer,\cite{krane_exchange_2023} a $21\%$ reduction in $U_{\mathrm{eff}}$ for 2T–H3T accounts for the enhanced spin coupling---highlighting the potential of wavefunction engineering to modulate exchange interactions in molecular spin systems.

\noindent
Controlled dehydrogenation of the H3T unit via tip-induced manipulation\cite{turco_observation_2023,zhao_tailoring_2024} yields the 2T--3T dimer (\textbf{1}), which serves as the ferrimagnetic unit for the trimeric structures.

The three resulting spin-carrying orbitals, $\psi_i$, are shown in Figure~\ref{fig: magnetic dimers}c. Notably, the 3T-localized orbital $\psi_3$ does not hybridize with $\psi_1$, justifying our spin-chain-like model (Figure~S1). The STS spectra in Figure~\ref{fig: magnetic dimers}c exhibit two symmetric steps with higher intensity at the 2T unit, and weaker features along with a zero-bias resonance at the 3T site. The latter is a hallmark of the degenerate doublet ground state of the 2T-3T system and its spatial distribution coincides with $\psi_3$, as evidenced by the constant-current \didv map on the right-hand side of Figure~\ref{fig:spin excitations}a-right. In contrast, the inelastic doublet–quartet spin excitation predominantly localizes at the 2T unit, as revealed in the left part of Figure~\ref{fig:spin excitations}a. Simulated spectra based on the corresponding Heisenberg model (black curves) reproduce the experimental data with an exchange coupling of $J_{2,3} = 54$ meV.

We note that both, experimental and theoretical spectra, recorded at the 2T unit reveal a characteristic zero-bias dip--—which is a spectroscopic signature of the ferromagnetic Kondo effect.\cite{turco_kondo_2026}

Having established the coupling constants $J_{2,\mathrm{H3}}$ and $J_{2,3}$, we now examine the trimeric structures 3T--2T--H3T, 3T--2T--3T, and 2T--3T--2T, corresponding to total spin ground states of $S = 1$, $S = \nicefrac{3}{2}$, and $S = 0$, respectively. 
The results summarized in Figure~\ref{fig:magnetic trimers} include a detailed low-bias STS analysis of the relevant magnetic excitations and a comparison with the corresponding simulated \didv curves.

The magnetic spectrum of 3T–2T–H3T (Figure~\ref{fig:magnetic trimers}a) shows inelastic transitions from the triplet ground state to singlet, triplet, and quintet states, which are well reproduced by the calculated \didv spectra using the previously determined exchange couplings $J_{2,\mathrm{H3}}$ and $J_{2,3}$.

Figures~\ref{fig:magnetic trimers}b and \ref{fig:magnetic trimers}c show low-bias STS data of 3T–2T–3T (\textbf{2}) and 2T–3T–2T (\textbf{3}), respectively. In 3T–2T–3T, asymmetric spin coupling yields an uncompensated spin-$\nicefrac{3}{2}$ ground state, with excitations to quartet and sextet states clearly resolved in the \didv{} spectra (green and purple traces). Constant-height \didv{} maps (Figure~\ref{fig:spin excitations}b) reveal the spatial distribution of these excitations: the quartet--sextet transition localizes on the 2T unit, while transitions to doublet and quartet states appear at the 3T sites. As in the 2T--3T dimer, the trimer exhibits signs of both ferromagnetic and overscreened behavior, evidenced by a zero-bias peak at the 3T sites and a dip at the 2T site (Figure~\ref{fig:magnetic trimers}b).\cite{turco_kondo_2026}

In contrast, the 2T-3T-2T trimer, with two 2T units coupled to a central 3T, is in a fully compensated singlet ground state ($S = 0$). STS spectra (Figure~\ref{fig:magnetic trimers}c) reveal two distinct inelastic excitations to triplet states, at 29 and 55~meV. Experimentally, both excitations appear at the 2T and 3T sites, whereas the Heisenberg model (see SI) predicts the 55~meV transition to be localized only at the 2T units. This discrepancy may arise from the simplified assumption that only one of the 3T's two degenerate zero modes couples to the neighboring 2T. While valid for dimers, this picture appears to break down in the symmetric 2T--3T--2T trimer, where both 2T units can hybridize with the 3T. This necessitates a more refined model of the coupling mechanism.\cite{mishra_observation_2021,catarina_hubbard_2022}

\section*{Conclusion}
\label{conclusion}

We have demonstrated an antiferromagnetic heterospin coupling motif as a robust strategy for engineering complex spin Hamiltonians in all-carbon systems. Through on-surface synthesis, we covalently couple $S = \nicefrac{1}{2}$ and $S = 1$ triangular nanographenes to construct three distinct ferrimagnetic configurations. Tip-induced dehydrogenation provides an additional tuning knob to tailor the magnetic properties of the resulting $\pi$-conjugated topologies, enabling access to all ground states from $S = 0$ to $S = \nicefrac{3}{2}$. High-resolution STS of the dimeric units, serving as elemental coupling motifs, yields Heisenberg parameters that accurately reproduce the magnetic excitations of the more complex trimer structures, validating the underlying spin model. The resulting spin Hamiltonians feature a rich manifold of spin multiplets and excitations, exemplifying prototypical multilevel quantum systems with tunable and well-defined spin states.

These results establish a bottom-up route to tailored spin architectures and provide a foundation for realizing 1D and 2D non-centrosymmetric lattices, where broken symmetry is predicted to stabilize ferrimagnetic ground states and correlated spin phases.\cite{catarina_broken-symmetry_2023,ortiz_theory_2022}

\section*{Acknowledgements}
This research was financially supported by the EU Graphene Flagship (Graphene Core 3, 881603), ERC Starting Grant (INSPIRAL, 716139), H2020-MSCA-ITN (ULTIMATE, No. 813036), Swiss National Science Foundation (SNF-PiMag, No. CRSII5\_205987 and 212875, PP00P2\_170534 and PP00P2\_198900), SNSF Consolidator Grant (CASCADER, TMCG-2\_213829), EIC-2022-Pathfinder Open (ATYPIQUAL, 101099098), the National Natural Science Foundation of China for funding (grant no. 92463307), and the Werner Siemens Foundation (CarboQuant).
E.T. would like to acknowledge Gon\c{c}alo Catarina for fruitful scientific discussions. Skillful technical assistance by Lukas Rotach is gratefully acknowledged.

\section*{Conflict of Interest}

The authors declare no conflict of interest.

\section*{Data Availability}
The raw NMR data are freely available on Zenodo at \url{https://zenodo.org/record/15603117} \\ (DOI: \href{https://doi.org/10.5281/zenodo.15603117}{10.5281/zenodo.15603117}).

%%%%%%%%%%%%%%%%%%%%%%%%%%%%%%%%%%%%%%%%%%%%%%%%%%%%%%%%%%
%%%%%%%%%%%%%%%%%%%%%%%%%%%%%%%%%%%%%%%%%%%%%%%%%%%%%%%%%%

%%%%%%%		References			%%%%%%% 

\setlength{\bibsep}{0.0cm}
\bibliographystyle{Wiley-chemistry}
\bibliography{references}

\clearpage

\end{document}

\maketitle

\vspace{12cm}
\section*{Data Availability}
The raw NMR data are freely available on Zenodo at \url{https://zenodo.org/record/15603117} \\ (DOI: \href{https://doi.org/10.5281/zenodo.15603117}{10.5281/zenodo.15603117}).

\vspace{0.5cm}
\newpage
\section{Experimental Methods}

\subsection{Sample Preparation and Scanning Probe Measurements}
\vspace{8pt}

STS measurements were performed with a commercial low-temperature STM/AFM from Scienta Omicron operated at a temperature of $4.5$ K and a base pressure below $5 \cdot 10^{-11}$ mbar. The Au(111) single crystal surfaces were prepared by iterative Ar$^+$ sputtering and annealing cycles. Before sublimation of molecules, the surface quality was verified through STM imaging. The powders of \textbf{7} and \textbf{11} precursors were filled into quartz crucibles of a home-built evaporator and sublimed at 210 ° C and 300 °C, respectively, on the surfaces of the single crystal. Instead, the molecular precursor \textbf{14} was sublimed via direct current heating of a Silicon wafer. STM images were acquired in both constant-current (overview and high-resolution imaging) and constant-height (bond-resolved imaging) modes, \didv spectra were acquired in constant-height mode, and \didv maps were acquired in constant-current mode. Indicated bias voltages are given with respect to the sample. Unless otherwise noted, all measurements were performed with metallic tips. Differential conductance \didv spectra and maps were obtained with a lock-in amplifier. Modulation voltages for each measurement are provided in the respective figure caption. Bond-resolved STM and \textit{nc}-AFM images were acquired in constant-height mode with CO-functionalized tips at low bias voltages while recording the current signal. Open feedback parameters on the molecular species and the subsequent lowering of the tip height ($\Delta z$) for each image are provided in the respective figure captions. The data was processed with Wavemetrics Igor Pro software.
\subsection{Hydrogen passivation }
In figure 3 of the main text, dihydro intermediates of \textbf{1} and \textbf{2} are reported. These structures are formed by passivation of the \textit{active} spin sites by atomic hydrogen diffusing on the metal surface, subsequently to the cyclodehydrogenation reactions. Therefore, it naturally occurs to find structures whose unpaired electrons are quenched by CH\textsubscript{2} groups, thus allowing the detection and sequential manipulation of intermediates with various spin ground states $S$. More details on the tip-based manipulation method can be found in Ref. \citenum{turco_observation_2023} and Ref. \citenum{zhao_tailoring_2024}.

\newpage

\section{Computational Methods}

\subsection{Tight-binding and mean-field Hubbard calculations}

TB-MFH calculations were performed by numerically solving the mean-field Hubbard Hamiltonian with third-nearest-neighbor hopping.

The corresponding Hamiltonian reads as
\begin{equation}
\mathcal{\hat{H}}_\mathrm{MFH} = \sum_{j} \sum_{\langle \alpha,\beta \rangle j,\sigma } t_j \hat{c}^\dagger_{\alpha,\sigma} \hat{c}_{\beta,\sigma} + U \sum_{\alpha, \sigma} \langle n_{\alpha,\sigma} \rangle n_{\alpha,\Bar{\sigma}}- U \sum_{\alpha} \langle n_{\alpha,\uparrow} \rangle \langle n_{\alpha,\downarrow} \rangle  ,
\end{equation}

Here, ${c_{\alpha,\sigma}}^\dagger$ and $c{_\beta,\sigma}$  denote the spin selective ($\sigma \in {\uparrow, \downarrow} $) creation and annihilation operator at sites $\alpha$ and $\beta$, $\langle\alpha,\beta\rangle_j$ ($j={1,3}$) denotes the nearest-neighbor and third-nearest-neighbor sites for j = 1, and 3, respectively, $t_j$ denotes the corresponding hopping parameters (with $t_1$  = 2.7 eV and $t_3  = 0.1 t_1$ for nearest-neighbor and third-nearest-neighbor hopping), U denotes the on-site Coulomb repulsion, $n_{\alpha,\sigma}$ denotes the number operator, and $\langle n_{\alpha,\sigma} \rangle$ denotes the mean occupation number at site $\alpha$. Orbital electron densities, $\rho$, of the $n^{th}$-eigenstate with energy $E_n$ have been simulated from the corresponding state vector $a_{n,i,\sigma}$ by

\begin{equation}
\rho_{n,\sigma}(\Vec{r}) = \Bigg| \sum_i a_{n,i,\sigma}\phi_{2p_z}(\Vec{r}-\vec{r_i}
)\bigg|^2  ,
\end{equation}

where i denotes the atomic site index and $\phi_{2p_z}$ denotes the Slater $2p_z$ orbital for carbon.
All TB-MFH calculations presented in the manuscript were done in the third-nearest-neighbor approximation and using an on-site Coulomb term $U = 1.2\, \lvert t_1 \rvert$.

\newpage
\subsection{Calculation of differential conductance spectra}
First, the spin Hamiltonian is constructed by a chain of spin-1/2 units, whereas each triangulene unit (3T) is considered to consist of two S=1/2 spins, coupled by a ferromagnetic Heisenberg-like exchange $J_\text{FM} = -1\,\text{eV}$. The phenalenyl radical (2T) and the hydrogenated triangulene (H3T) are represented by a single spin-1/2. The units are coupled by one of the two coupling parameters $J_{2,3}=54\,\text{meV}$ between 2T and 3T and $J_{2,H3}=60\,\text{meV}$ between 2T and H3T.

The d\textit{I}/d\textit{V} spectra were simulated by introducing a perturbative spin scattering term into the Hamiltonian, which accounts for spin-flip processes up to third order in the interaction matrix elements \cite{ternes_spin_2015}.

\begin{figure}[h!]
    \centering
    \includegraphics[width=0.5\linewidth]{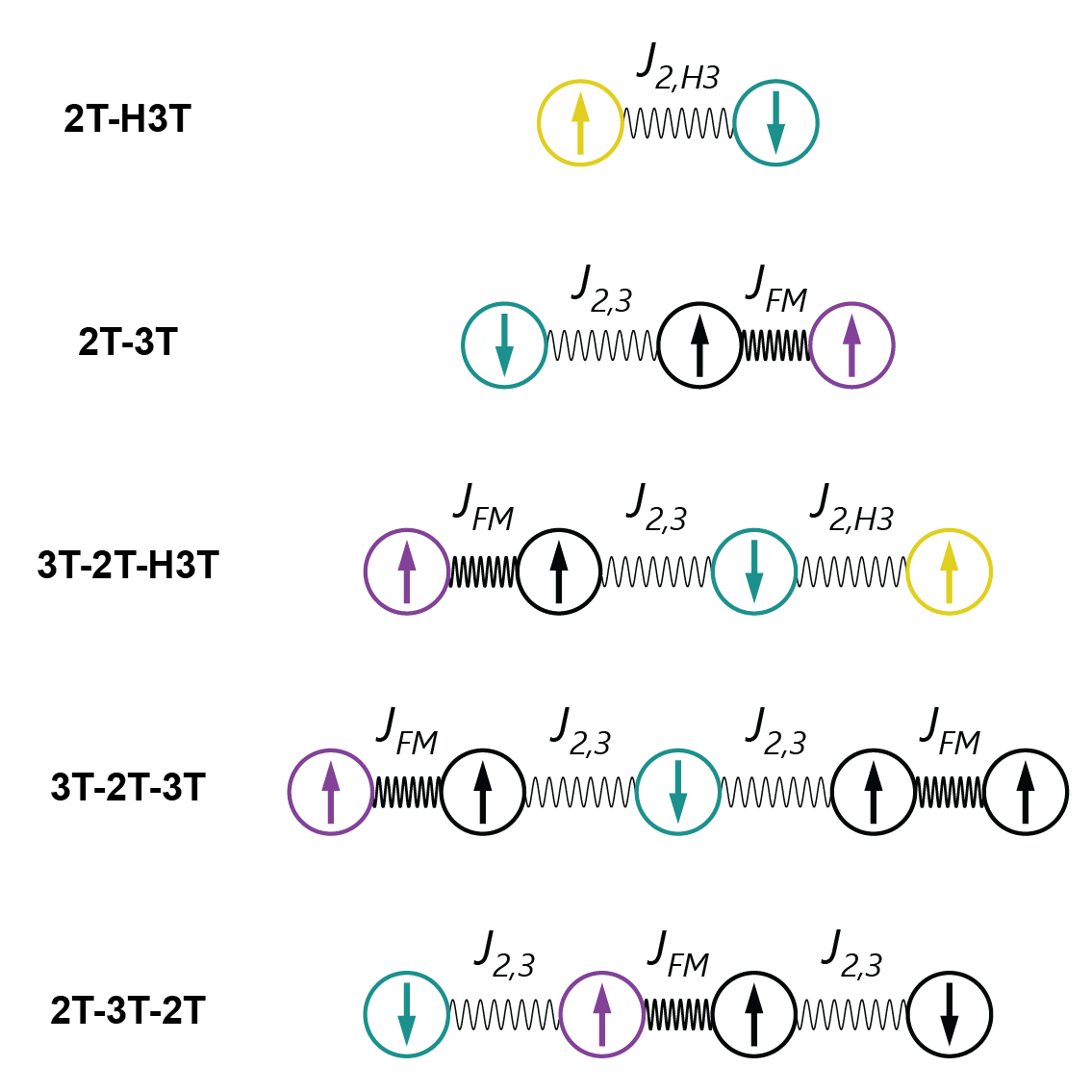}
    \caption{Sketch of Heisenberg models used for simulation of differential conductance spectra.
    Each unit represents an S=1/2 spin, coupled by one of three different coupling strengths to the neighbor.
    The colored units indicate the considered spins for the respective spectra, shown in Figures 4 and 6 of the main text.}
    \label{fig:spin_model}
\end{figure}

\begin{table}[]
    \renewcommand{\arraystretch}{1.2}
    \centering
    \begin{tabular}{c||c|c||c|c||c|c|c||c|c||c|c||}
 & \multicolumn{2}{c||}{2T-H3T} & \multicolumn{2}{c||}{2T-3T} &
 \multicolumn{3}{c||}{3T-2T-H3T} & \multicolumn{2}{c||}{3T-2T-3T} &
 \multicolumn{2}{c||}{2T-3T-2T} \\
Parameter & H3T & 2T & 2T & 3T & 3T & 2T & H3T & 3T & 2T & 2T & 3T \\
\hline
$U$ & -0.1 & -0.1 & -0.08 & 0 & 0 & -0.1 & -0.1 & 0 & -0.2 & -0.1 & -0.1 \\
$J\rho$ & -0.3 & -0.3 & -0.25 & -0.3 & -0.2 & -0.4 & -0.4 & -0.2 & -0.3 & -0.2 & -0.2 \\
T & 30 & 30 & 20 & 15 & 15 & 25 & 15 & 15 & 30 & 15 & 15 \\
%A & 0.3 & 0.37 & 0.5 & 0.4 & 0.6 & 0.35 & 0.6 & 1.5 & 1.5 & 0.35 & 0.35 \\
$x_0$ & -1.15 & -1.15 & -1.15 & -0.15 & -1.15 & -1.15 & -1.15 & -1.15 & -1.15 & -1.15 & -1.15 \\
slope & -0.1 & -0.1 & -0.1 & -0.4 & 0 & 0.4 & 0.2 & -0.4 & 0.2 & -0.3 & -0.3 \\
%spin & 0 & 1 & 0 & 2 & 0 & 2 & 3 & 0 & 2 & 0 & 1
    \end{tabular}
    \caption{}
    \label{tab:simulations}
\end{table}

Table~\ref{tab:simulations} summarizes the parameter used to simulate the d\textit{I}/d\textit{V} shown in the main text: $U$ is the asymmetry parameter for broken electron--hole symmetry, $J\rho$ the coupling strength to the substrate, $T$ the temperature in Kelvin and $x_0$ the x-offset in meV (see Ref. \citenum{ternes_spin_2015}).

\section{Additional experimental data}

\begin{figure}[h!]
    \centering
    \includegraphics[width=0.7\linewidth]{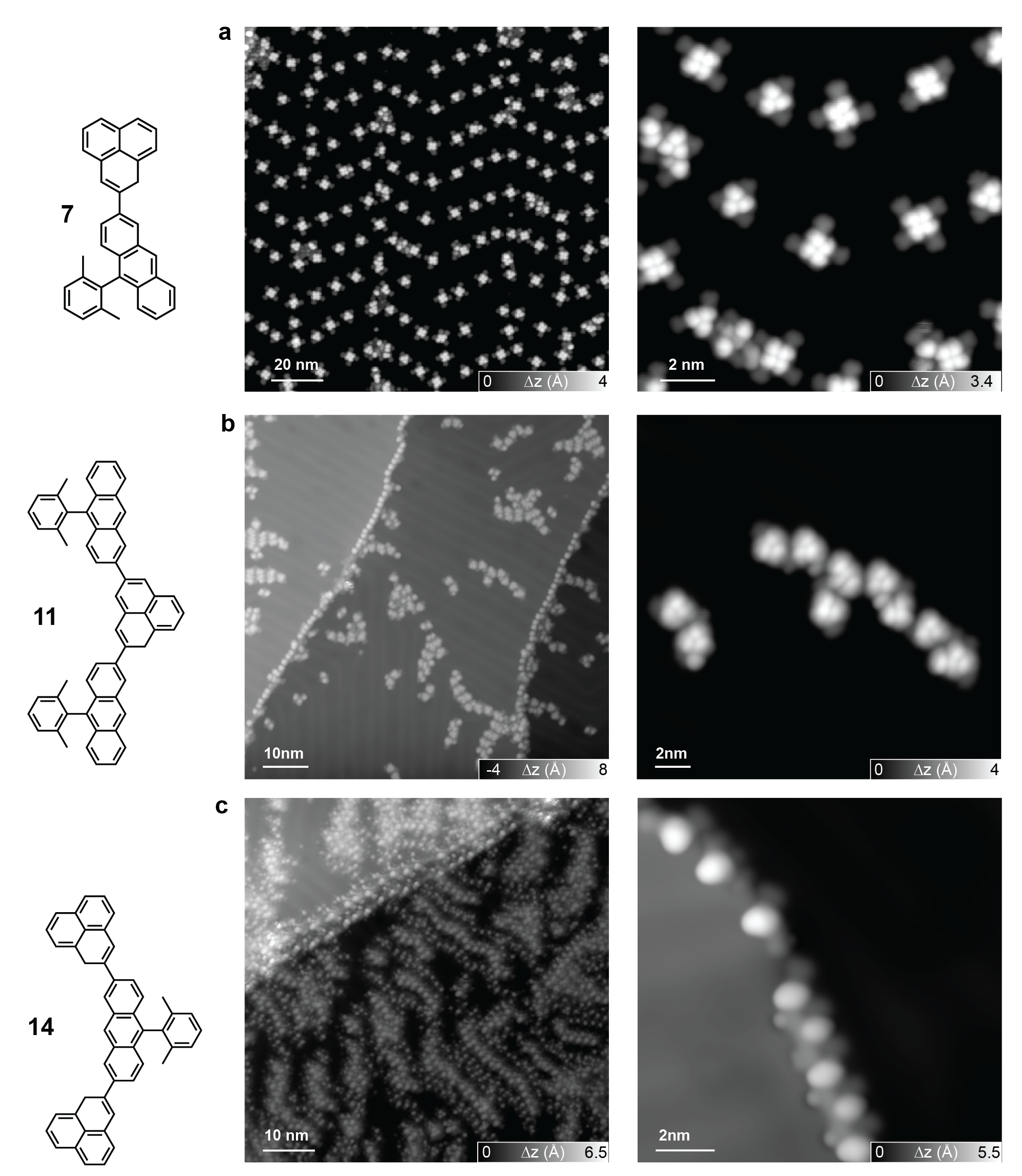}
    \caption{Molecular precursors deposited on Au(111) at room temperature. From left to right: chemical structures of precursors \textbf{7}, \textbf{11}, and \textbf{14}; (a–c) overview STM images showing submonolayer coverage; high-resolution STM images (right) of the corresponding self-assembled molecular clusters. Tunneling parameters: -1 V/30 pA (a,c-left), -1.5 V/30 pA (b), (c-right) -0.4 V/50 pA. The ambiguous appearance of (c-left) is due to a double-tip effect.  }
    \label{fig:RT phase}
\end{figure}

\begin{figure}
    \centering
    \includegraphics[width=\linewidth]{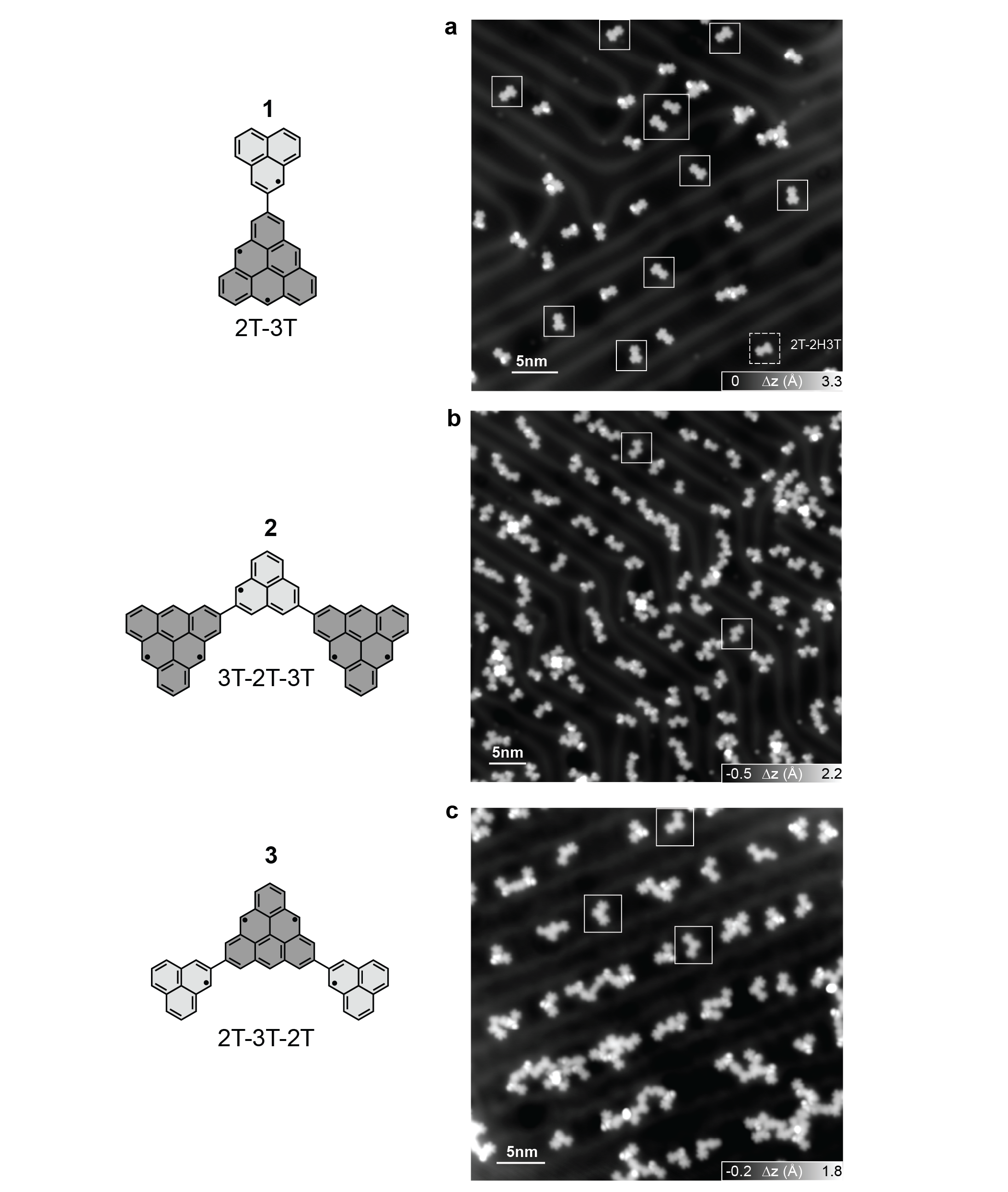}
    \caption{Thermally assisted on-surface synthesis towards target compounds \textbf{1-3}. Starting from the submonolayer coverage of the molecular precursors (described in Figure \ref{fig:RT phase}), each sample was annealed to 320°C for few minutes to trigger the oxidative ring closure of the methyl groups. (a,b,c) STM images acquired subsequently to the annealing step, starting from precursors \textbf{7}, \textbf{11} and \textbf{14}. In each STM image, the corresponding target structures are highlighted with white frames. Notably, the yield of target molecules is higher for \textbf{1} than for \textbf{2} and \textbf{3}. Along with the target structures we always observe a coexistence of covalently coupled molecular clusters and unreacted molecular precursors. Tunneling parameters: -1 V/50 pA (a,b), -0.1 V/50 pA (c).}
    \label{fig:annealed phase}
\end{figure}

\begin{figure}
    \centering
    \includegraphics[width=\linewidth]{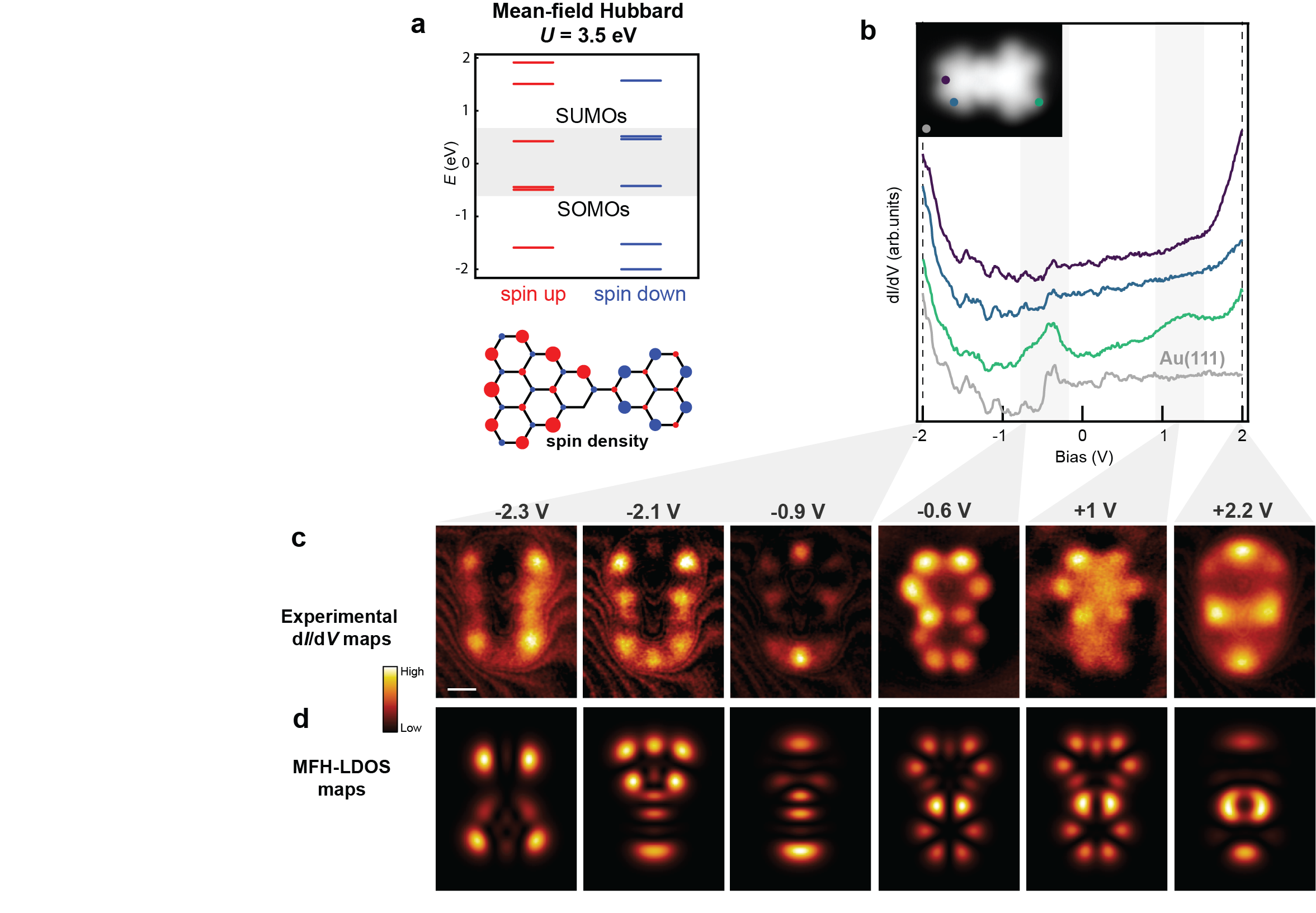}
    \caption{Electronic characterization of 2T-3T(\textbf{1}) on Au(111). (a) MFH energy spectrum (top) and spin polarization plot (bottom), where blue and red filled circles denote mean populations of spin up and spin down electrons (circles size scaling with the absolute value), respectively. Spin-carrying orbitals are highlighted by shaded gray overlays. (b) \didv spectroscopy on \textbf{1} revealing molecular orbital resonances (open feedback parameters: $V$=–2.0 V, $I$= 350 pA; \LI = 20 mV). Acquisition positions are indicated in the HR-STM image shown as inset ($V$ = –0.6 V, $I$= 150 pA).(c) Constant-current \didv maps of the molecular orbital resonances, acquired with a metal tip. All the \didv maps were acquired with a lock-in modulation \LI = 20 mV and a current setpoint $I=400$ pA. (d) MFH-TB LDOS of HOMO-3, HOMO-2, HOMO-1, SOMOs, SUMOs, LUMO+1, respectively.}
    \label{fig:STS 2T-3T}
\end{figure}

\begin{figure}
    \centering
    \includegraphics[width=0.5\linewidth]{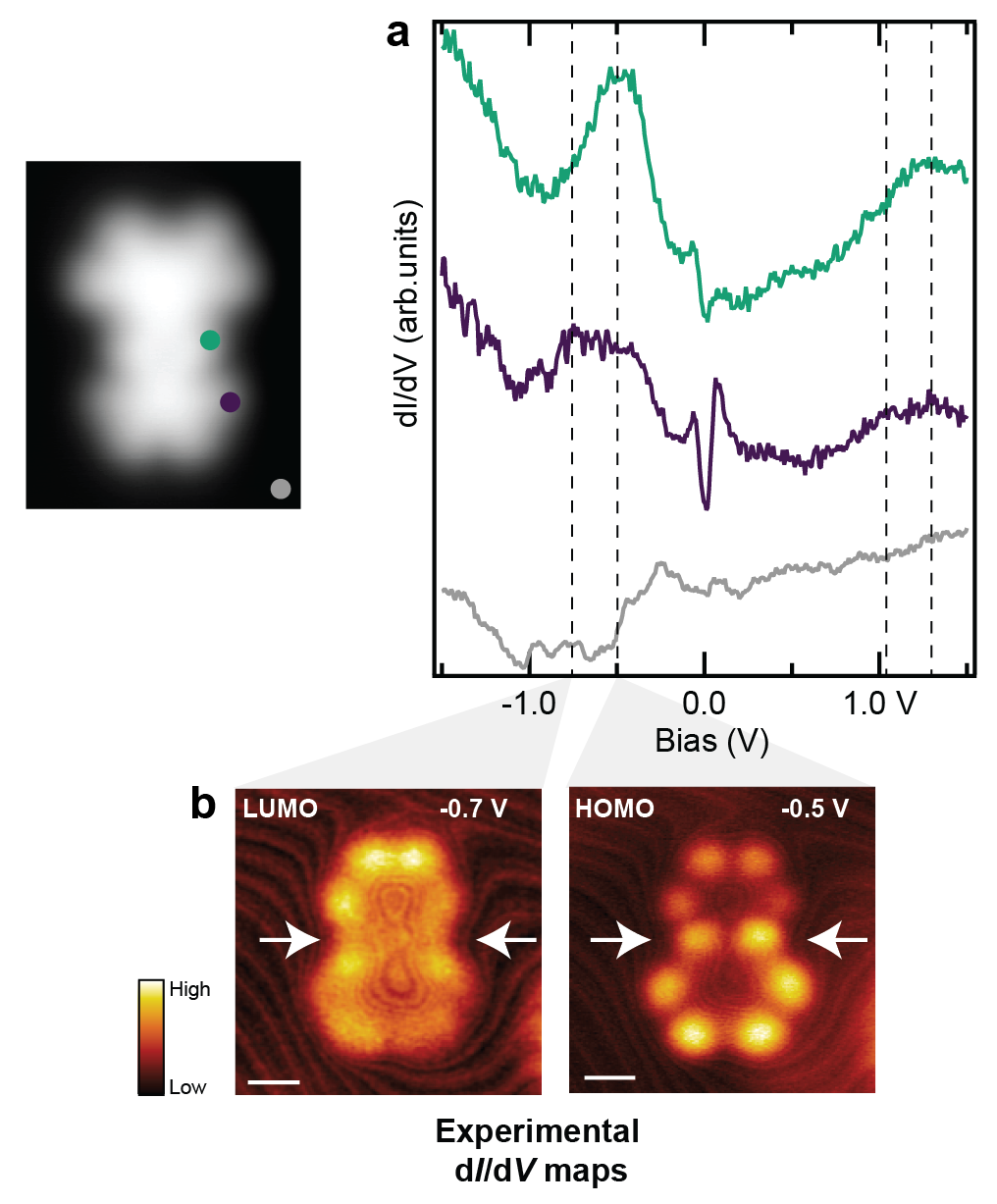}
   \caption{
(a) \didv spectra acquired with a CO-functionalized tip on \textbf{1} adsorbed on Au(111). Vertical dotted lines indicate the energies of the HOMO and LUMO resonances for both polarities\cite{krane_exchange_2023}. The corresponding measurement positions are marked by circles in in the STM ($V$ = –0.6 V, $I$= 150 pA) image on the left . Spectra were recorded with the feedback loop opened at $-1.5$~V, 300~pA, and \LI = 20~mV.  
(b) \didv maps recorded at negative sample biases with a metal tip, reflecting the spatial distribution of the LUMO (left) and HOMO (right). Imaging parameters: 300~pA, \LI =  20~mV. For positive biases, the broadening of the molecular orbital resonances hinders the visualization of HOMO/LUMO as distinct contributions. These observations are in line with what has been shown in Ref. \cite{krane_exchange_2023}. 
}
    \label{fig:many-body}
\end{figure}

\begin{figure}[h!]
    \centering
    \includegraphics[width=0.5\linewidth]{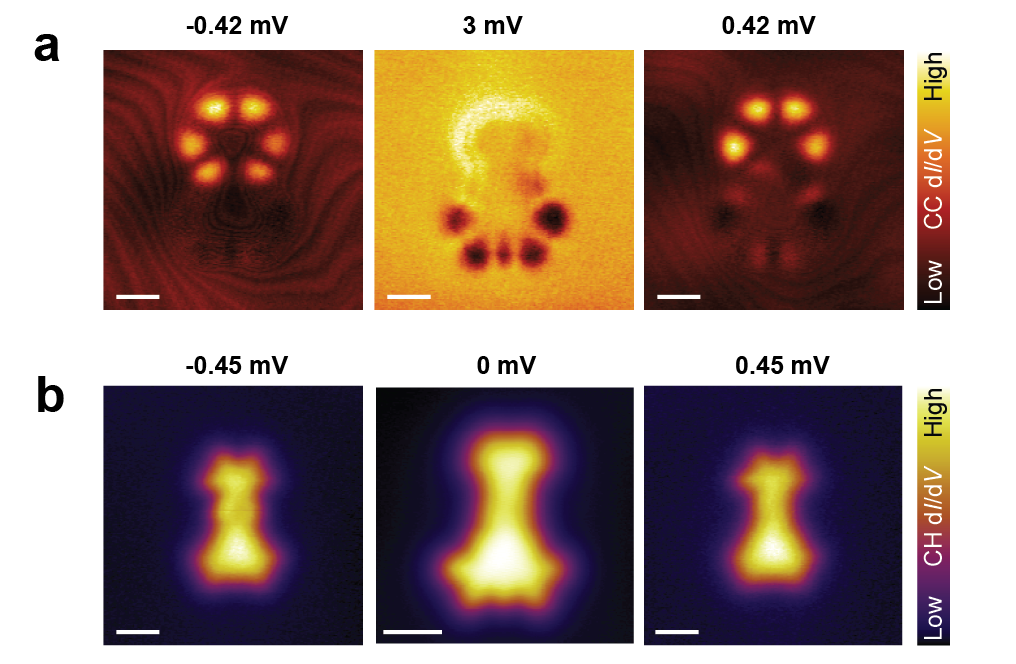}
    \caption{Constant-current and constant-height spin excitation maps of 2T-3T(\textbf{1}). (a) Constant-current \didv maps of the doublet-quartet spin excitation (left/right) and Kondo-like resonance (center). Current setpoints: (a)-left/right 1 nA, (a)-center 100 pA. (b) Constant-height \didv maps of the doublet-quartet spin excitation (left/right) and Kondo-like resonance (center). Feedback opened on the molecule at: $\pm$45 mV/1 nA (left/right), -15 mV/800 pA (center). The Lock-in modulation \LI used is 2 mV for all maps except the constant-height map at 0 mV, which was acquired with \LI= 4 mV. Scale bars (a,b): 0.5 nm.   }
    \label{fig:Suppl spin excitation maps 2T-3T}
\end{figure}

\begin{figure}[h!]
    \centering
    \includegraphics[width=.8\linewidth]{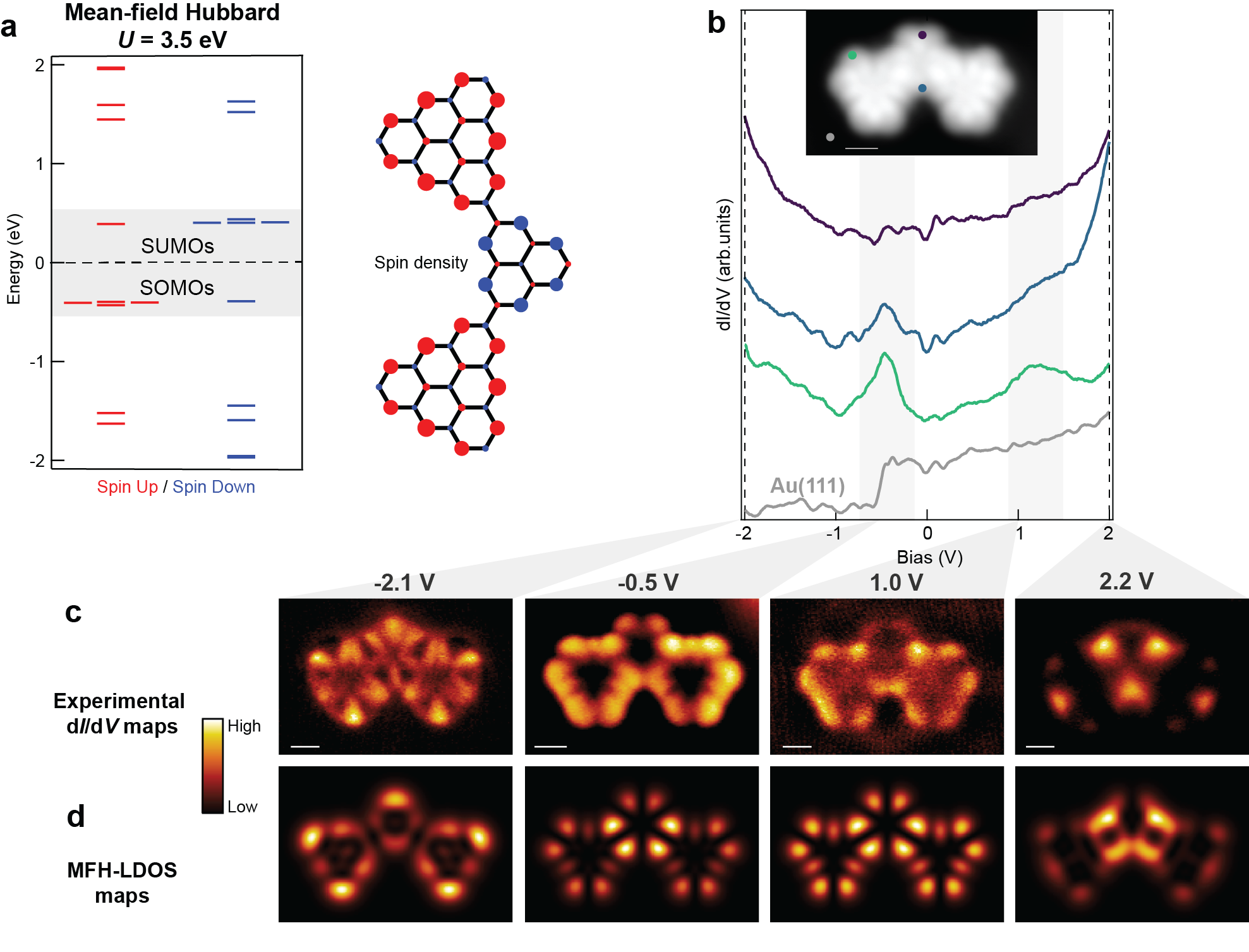}
    \caption{Electronic characterization of 3T-2T-3T(\textbf{2}) on Au(111).(a) MFH energy spectrum (left) and spin polarization plot (right), where blue and red filled circles denote mean populations of spin up and spin down electrons (circles size scaling with the absolute value), respectively. Spin-carrying orbitals are highlighted by shaded gray overlays. (b) \didv spectroscopy on \textbf{2} revealing molecular orbital resonances (open feedback parameters: $V$=–2.0 V, $I$= 300 pA; \LI = 30 mV). Acquisition positions are indicated in the HR-STM image shown as inset ($V$ = –0.1 V, $I$= 150 pA).(c) Constant-current \didv maps of the molecular orbital resonances, acquired with a CO-functionalized tip. All the \didv maps were acquired with a lock-in modulation \LI = 30 mV and a current setpoint $I=250$ pA. (d) MFH-TB LDOS of HOMO-1, SOMOs, SUMOs, LUMO+1, respectively.}
    \label{fig:STS 3T-2T-3T}
\end{figure}

\begin{figure}[h!]
    \centering
    \includegraphics[width=.8\linewidth]{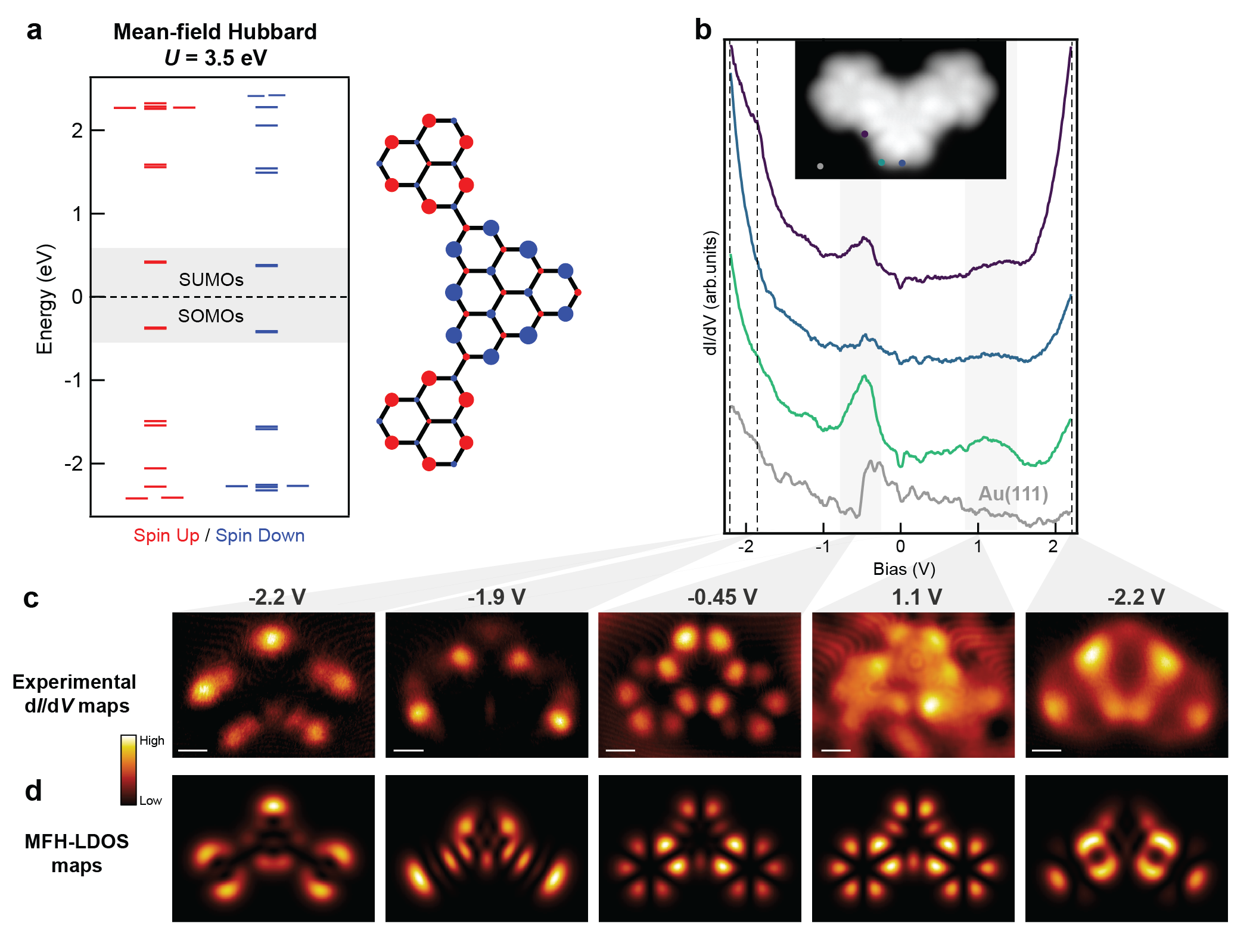}
    \caption{Electronic characterization of 2T-3T-2T (\textbf{3}) on Au(111). (a) MFH energy spectrum (left) and spin polarization plot (right), where blue and red filled circles denote mean populations of spin up and spin down electrons (circles size scaling with the absolute value), respectively. Spin-carrying orbitals are highlighted by shaded gray overlays. (b) \didv spectroscopy on \textbf{3} revealing molecular orbital resonances (open feedback parameters: $V$=–2.2 V, $I$= 300 pA; \LI = 30 mV). Acquisition positions are indicated in the HR-STM image shown as inset ($V$ = –0.1 V, $I$= 100 pA).(c) Constant-current \didv maps of the molecular orbital resonances, acquired with a metal tip. All the \didv maps were acquired with a lock-in modulation \LI = 30 mV and a current setpoint $I=350$ pA. (d) MFH-TB LDOS of HOMO-2, HOMO-1, SOMOs, SUMOs, LUMO+1, respectively.}
    \label{fig:STS 2T-3T-2T}
\end{figure}
\newpage

\clearpage
\FloatBarrier

\bibliographystyle{Wiley-chemistry}
\bibliography{SI_references}



\section*{Supplementary material}

# Details of the In-Solution Synthesis of Precursors

## 4.1 General Methods and Materials

Unless otherwise specified, commercially available starting materials, reagents, catalysts, and dry solvents were used without further purification. All reactions were carried out under an inert atmosphere using standard Schlenk and vacuum-line techniques. Starting materials were sourced from TCI, Sigma-Aldrich, abcr, Alfa Aesar, Acros Organics, and VWR. Catalysts were obtained from Strem Chemicals.

Column chromatography was performed using silica gel ($SiO_2$, particle size 0.063–0.200 mm, VWR). Thin-layer chromatography (TLC) was conducted on silica-coated aluminum sheets with a fluorescence indicator (TLC silica gel 60 F254, Merck KGaA). Further purification by recycling gel permeation chromatography (r-GPC) was conducted on a JAI HPLC LC-9110II NEXT system equipped with a fraction collector (FC-3310) and connected in series with JAIGEL-1HH, JAIGEL-2HH, and JAIGEL-2.5HH columns, using HPLC-grade chloroform as the eluent.

NMR spectra were recorded on a Bruker AV-II 300 spectrometer (300 MHz for $^1$H, 76 MHz for $^{13}$C) and a Bruker FT-NMR Avance III 400 spectrometer (400 MHz for $^1$H and 101 MHz for $^{13}$C) and Avance 600 at 298 K unless indicated otherwise. Chemical shifts ($\delta$) are reported in parts per million (ppm) relative to the residual solvent peak (dichloromethane-$d_2$: $\delta(^1$H$) = 5.32$ ppm, $\delta(^{13}$C$) = 53.84$ ppm; chloroform-$d$: $\delta(^1$H$) = 7.26$ ppm, $\delta(^{13}$C$) = 77.16$ ppm). Coupling constants ($J$) are given in Hz. Peak multiplicities are denoted as follows: s = singlet, d = doublet, t = triplet, q = quartet, m = multiplet. Dichlormethane-$d_2$ ($\geq$99.5 atom% D) and 1,1,2,2-tetrachloroethane-$d_2$ ($\geq$99.5 atom% D) were purchased from Sigma-Aldrich.

Mass spectra were recorded with a gas chromatography mass spectrometry setup consisting of a Finnigan TraceGC Ultra and Thermo Electron Corporation Trace DSQ. High-resolution mass spectra were measured as EI (ThermoFischer Scientifics DFS), ESI, or APCI (ThermoFischer Scientifics QExactive) MS.

MALDI-TOF mass spectra were acquired on a Bruker Autoflex Speed MALDI-TOF mass spectrometer (Bruker Daltonics, Bremen, Germany). Samples for MALDI-TOF analysis were prepared by co-crystallizing the analyte with the matrix—either 1,8-dihydroxyanthracen-9(10*H*)-one (dithranol, > 98% purity, Fluka Analytical) or *trans*-2-[3-(4-*tert*-butylphenyl)-2-methyl-2-propenylidene]malononitrile (DCTB, >99% purity, Sigma-Aldrich)—in the solid state.

The key building blocks 2-bromo-10-(2,6-dimethylphenyl)anthracene (**4**),[5] 2,7-dibromo-10-(2,6-dimethylphenyl)anthracene (**12**),[5] and 2-bromo-1*H*-phenalen-1-one (**S1**)[6] were synthesized according to previously reported procedures.

## 4.2 Detailed Synthetic Procedures
### Synthesis of 2-(4,4,5,5-tetramethyl-1,3,2-dioxaborolan-2-yl)-1*H*-phenalen-1-one (5)

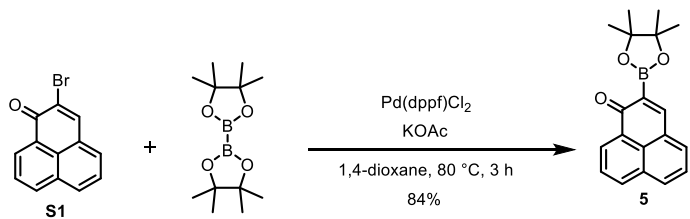

Compound **5** was prepared using a literature procedure;[7] the reaction was carried out under inert conditions; solvents were deoxygenated by bubbling with nitrogen while sonication in an ultrasonic bath for 15–30 min. In a sealed tube, bis(pinacolato)diboron (343 mg, 1.35 mmol), **S1** (350 mg, 1.35 mmol), potassium acetate (265 mg, 2.70 mmol), and Pd(dppf)Cl$_2$ (49 mg, 0.067 mmol, 5 mol%) were suspended in 1,4-dioxane (17 mL). The mixture was heated at 80 °C for 3 h. After cooling to room temperature, the crude mixture was diluted with water and extracted with methyl *tert*-butyl ether (MTBE). The combined organic layers were washed with water, followed by brine and dried over MgSO$_4$. After reducing the solvent under reduced pressure, the crude product was passed through a pad of silica eluding with MTBE. Evaporation of the solvent afforded the desired compound as a brown resin (0.49 g, 0.95 mmol, 84%). Purity: 70% by $^1$H-NMR; the desired compound could not be separated from the residual pinacolborane without decomposition. The mixture was used for cross-coupling reaction without further purification. Only a small fraction of the desired compound was purified (flash column chromatography (SiO$_2$, cyclohexane/ethyl acetate = 10:1 v/v) for the characterization of the compound.

HRMS (EI) *m/z*: [*M*]$^+$ calcd. for C$_{19}$H$_{19}$BO$_3$: 306.1422, found: 306.1430. $^1$H NMR (400 MHz, CDCl$_3$): $\delta$ 8.62 (dd, *J* = 7.3, 1.2 Hz, 1H), 8.32 (s, 1H), 8.16 (dd, *J* = 8.1, 1.2 Hz, 1H), 8.04 (d, *J* = 8.6 Hz, 1H), 7.81–7.71 (m, 2H), 7.59 (dd, *J* = 8.3, 7.0 Hz, 1H), 1.40 (s, 12H). $^{13}$C NMR (101 MHz, CDCl$_3$): $\delta$ 187.1, 152.2, 134.3, 132.6, 132.2, 132.0, 130.8, 130.0, 129.9, 128.5, 127.8, 127.5, 126.7, 84.1, 25.0.

### Synthesis of 2-(10-(2,6-dimethylphenyl)anthracen-2-yl)-1*H*-phenalen-1-one (6)

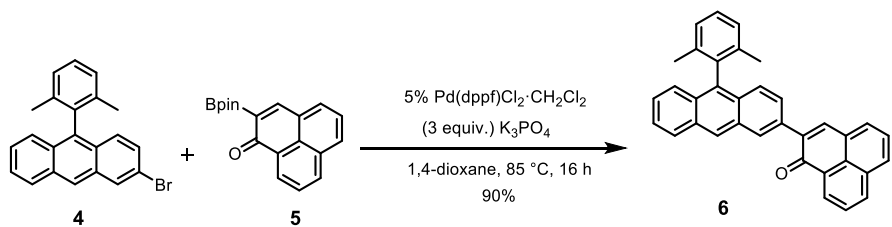

In a 25 mL Schlenk tube, compound **4** (47.2 mg, 0.131 mmol), compound **5** (20.0 mg, 65.3 µmol), Pd(dppf)Cl$_2$·CH$_2$Cl$_2$ (2.7 mg, 3.3 µmol), and K$_3$PO$_4$ (42 mg, 0.20 mmol) were added. The tube was evacuated and backfilled with argon three times. 1,4-Dioxane (5 mL) was then added, and the reaction mixture was stirred at 85 °C for 16 h under an argon atmosphere. After cooling to room temperature, the reaction mixture was poured into water and extracted with dichloromethane (DCM) three times. The combined organic layers were washed with brine, dried over anhydrous MgSO$_4$, and concentrated under reduced pressure. The crude product was purified by silica gel column chromatography (iso-hexane/DCM = 1:1 v/v) to afford compound **6** as an orange solid (27 mg, 90% yield).

HR-MALDI-TOF MS (matrix: DCTB) *m/z*: [*M*]$^+$ calcd. for C$_{35}$H$_{24}$O: 460.1822, found: 460.1801. $^1$H NMR (300 MHz, dichlormethane-*d*$_2$): $\delta$ 8.72 (dd, *J* = 7.4, 1.2 Hz, 1H), 8.61 (s, 1H), 8.47 (d, *J* = 1.8 Hz, 1H), 8.29 (dd, *J* = 8.2, 1.2 Hz, 1H), 8.15–8.05 (m, 3H), 7.94–7.82 (m, 2H), 7.73–7.63 (m, 2H), 7.55–7.27 (m, 7H), 1.78 (s, 6H). $^{13}$C NMR (76 MHz, dichlormethane-*d*$_2$): $\delta$ 184.5, 140.5, 139.4, 138.3, 138.0, 136.0, 135.3, 134.4, 132.7, 132.5, 132.2, 132.11, 132.07, 131.3, 130.6, 130.4, 129.5, 129.33, 129.28, 128.7, 128.3, 128.0, 127.9, 127.7, 127.6, 127.4, 126.5, 126.3, 125.9, 125.8, 20.3 (one carbon resonance could not be resolved because of signal overlap).

### Synthesis of 10-(2,6-dimethylphenyl)-2-(1*H*-phenalen-2-yl)anthracene (7) (Mixture of isomers)

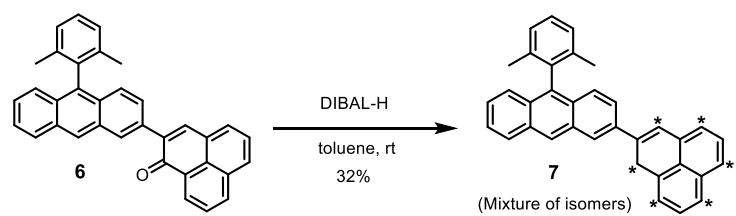

In a 10 mL Schlenk tube, compound **6** (13 mg, 28 μmol) was dissolved in deoxygenated toluene (1 mL). Diisobutylaluminum hydride (DIBAL-H, 56 uL, 1 M in hexane, 2 equiv.) was added dropwise at room temperature. Upon addition, the reaction mixture immediately changed color from orange to light yellow and exhibited sky-blue fluorescence under 365 nm UV light. The reaction was stirred at room temperature overnight. Then, the reaction mixture was passed through a plug of Florisil (6 × 2 cm) using deoxygenated toluene as the eluent. The solvent was removed under reduced pressure to yield the crude product. Thin-layer chromatography (TLC) revealed three fluorescent spots, all exhibiting sky-blue emission under UV light. Further purification by recycling gel permeation chromatography (r-GPC, fraction collected at t = 80–84 min) afforded a light-yellow solid **7** (4 mg, 32% yield) as a mixture of isomers. NMR: Compound **7** is obtained as a mixture of isomers where the methylene groups can be shifted to any α-position (asterisks) of the phenalenyl subunit. The mixture of isomers gives a complex proton NMR spectrum with characteristic singlets for the methylene groups. The ratio between methylene groups, aromatic protons, and methyl groups is 2:18:6 matching the expected structures. High-resolution MALDI-TOF mass spectrometry (HR-MALDI-TOF MS) of the isomeric mixture showed a single dominant peak corresponding to $[M - H]^+$ at $m/z$ = 445.1927, in excellent agreement with the calculated value ($m/z$ = 445.1945, Δ = –4.0 ppm), and the observed isotopic pattern matched the simulated spectrum (Figures S16), confirming the molecular integrity of the compound. This isomeric mixture (**7**) was directly used as the precursor for the on-surface synthesis.

### Synthesis of 2-(10-(2,6-dimethylphenyl)anthracen-2-yl)-4,4,5,5-tetramethyl-1,3,2-dioxaborolane (8)

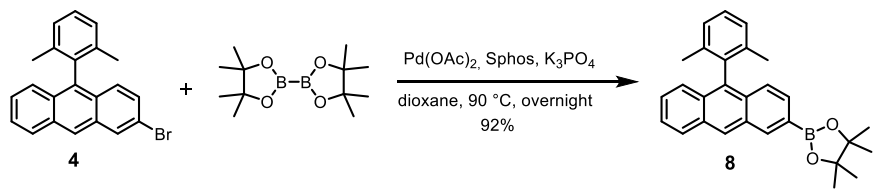

To a 50 mL Schlenk flask equipped with a magnetic stir bar, compound **4** (0.15 g, 0.40 mmol), 4,4,4′,4′,5,5,5′,5′-octamethyl-2,2′-bi(1,3,2-dioxaborolane) (0.20 g, 0.80 mmol), Pd(OAc)$_2$ (18 mg, 0.080 mmol, 10 mol%), SPhos (66 mg, 0.16 mmol), and K$_3$PO$_4$ (511 mg, 2.41 mmol) were added. The flask was evacuated and backfilled with argon three times. After further flushing with argon for 30 min, deoxygenated 1,4-dioxane (12 mL) was added. The reaction mixture was stirred at 90 °C overnight under an argon atmosphere. Upon completion, the reaction mixture was cooled to room temperature, poured into water, and extracted with ethyl acetate. The combined organic layers were dried over anhydrous MgSO$_4$, filtered, and concentrated under reduced pressure. The crude product was purified by silica gel column chromatography (iso-hexane/DCM = 1:1 v/v) to afford compound **8** as a light-yellow solid (150 mg, 92% yield).

HR-MALDI-TOF MS (matrix: dithranol) $m/z$: $[M]^+$ calcd. for C$_{28}$H$_{29}$BO$_2$: 408.2255, found: 408.2248. $^1$H NMR (300 MHz, dichloromethane-$d_2$): $\delta$ 8.59 (d, $J$ = 6.9 Hz, 2H), 8.14–8.08 (m, 1H), 7.62 (dd, $J$ = 8.7, 1.2 Hz, 1H), 7.52–7.32 (m, 5H), 7.31–7.24 (m, 2H), 1.72 (s, 6H), 1.39 (s, 12H). $^{13}$C NMR (76 MHz, dichloromethane-$d_2$): $\delta$ 138.3, 138.2, 138.0, 135.9, 132.2, 131.6, 131.1, 130.87, 130.1, 129.5, 128.3, 128.04, 127.95, 126.8, 126.2, 125.8, 125.2, 84.5, 25.3, 20.3 (one carbon resonance (presumably C–B) could not be detected).

**Synthesis of 2,5-dibromo-1*H*-phenalen-1-one (9)**

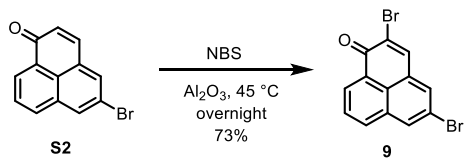

The compound was prepared using a modified literature procedure.[8] 5-Bromo-1*H*-phenalen-1-one (**S2**; 61 mg, 0.23 mmol), NBS (71 mg, 0.40 mmol), and neutral Al₂O₃ (327 mg) were ground in a mortar to obtain an evenly blended mixture which was placed in a test tube. The reaction mixture was heated at 45 °C overnight. Then, the solid mixture was extracted with CH₂Cl₂. After evaporation of the solvent, the crude product was purified by flash column chromatography (SiO₂, CH₂Cl₂) to afford the desired compound **9** as a yellow solid (58 mg, 0.17 mmol, 73%). HRMS (EI) *m/z*: [*M*]⁺ calcd. for C₁₃H₆Br₂O: 335.8780, found: 335.8780. ¹H NMR (400 MHz, CDCl₃): δ 8.69 (dd, *J* = 7.5, 1.2 Hz, 1H), 8.21 (d, *J* = 1.8 Hz, 1H), 8.18–8.11 (m, 2H), 7.86–7.77 (m, 2H). ¹³C NMR (101 MHz, CDCl₃): δ 178.4, 141.8, 134.7, 133.73, 133.71, 133.3, 132.5, 129.6, 128.7, 128.6, 127.4, 125.4, 120.7.

**Synthesis of 2,5-bis(10-(2,6-dimethylphenyl)anthracen-2-yl)-1*H*-phenalen-1-one (10)**

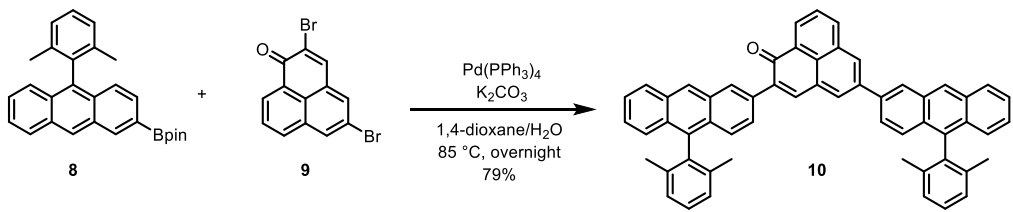

The compound was prepared using a modified literature procedure;[9] the reaction was carried out under inert conditions; solvents were deoxygenated by bubbling with nitrogen while sonication in an ultrasonic bath for 15–30 min.    2-(10-(2,6-Dimethylphenyl)anthracen-2-yl)-4,4,5,5-tetramethyl-1,3,2-dioxaborolane (**8**;   53 mg, 0.13 mmol), 2,5-dibromo-1*H*-phenalen-1-one (**9**; 22 mg, 0.060 mmol), K₂CO₃ (36 mg, 0.26 mmol), and Pd(PPh₃)₄ (7.4 mg, 6.0 µmol, 10 mol%) were suspended in 1,4-dioxane (7 mL) and water (1.75 mL). The mixture was heated at 85 °C overnight. After cooling to room temperature, the crude mixture was poured into water and extracted with CH₂Cl₂. The combined organic layers were washed with water, followed by brine, and dried over MgSO₄. After evaporation of the solvent, the crude product was purified by flash column chromatography (SiO₂, cyclohexane/ethyl acetate = 4:1 v/v) to afford the desired compound as an orange solid (37 mg, 0.050 mmol, 79%). HRMS (EI) *m/z*: [*M*]⁺ calcd. for C₅₇H₄₀O: 740.3074, found: 740.3086. ¹H NMR (400 MHz, CDCl₃): δ 8.77 (dd, *J* = 7.4, 1.2 Hz, 1H), 8.62 (s, 1H), 8.58 (s, 1H), 8.51 (d, *J* = 1.7 Hz, 1H), 8.45 (d, *J* = 1.9 Hz, 1H), 8.39 (d, *J* = 1.6 Hz, 1H), 8.34 (dd, *J* = 8.2, 1.2 Hz, 1H), 8.29 (d, *J* = 1.7 Hz, 1H), 8.15 (s, 1H), 8.13–8.06 (m, 2H), 7.89 (dd, *J* = 7.7 Hz, 1H), 7.77 (dd, *J* = 9.1, 1.9 Hz, 1H), 7.69 (dd, *J* = 9.0, 1.8 Hz, 1H), 7.64 (d, *J* = 9.0 Hz, 1H), 7.57–7.45 (m, 5H), 7.43–7.33 (m, 4H), 7.33–7.27 (m, 4H), 1.81 (s, 6H), 1.80 (s, 6H). ¹³C NMR (101 MHz, CDCl₃): δ 184.5, 140.0, 139.9, 139.6, 138.0, 137.9, 137.6, 137.4, 136.3, 136.0, 135.6, 135.1, 133.6, 132.7, 132.4, 132.0, 131.8, 131.7, 131.2, 131.0, 130.2, 130.1, 129.28, 129.26, 129.2, 129.0, 128.92, 128.85, 128.01, 127.99, 127.9, 127.7, 127.6, 127.3,

127.1, 127.04, 126.96, 126.6, 126.2, 126.12, 126.05, 126.0, 125.8, 125.42, 125.35, 20.29, 20.27 (four carbon resonances could not be resolved because of signal overlap).

**Synthesis of 3,3'-(1*H*-phenalene-2,5-diyl)bis(9-(2,6-dimethylphenyl)anthracene) (11)** (Mixture of isomers)

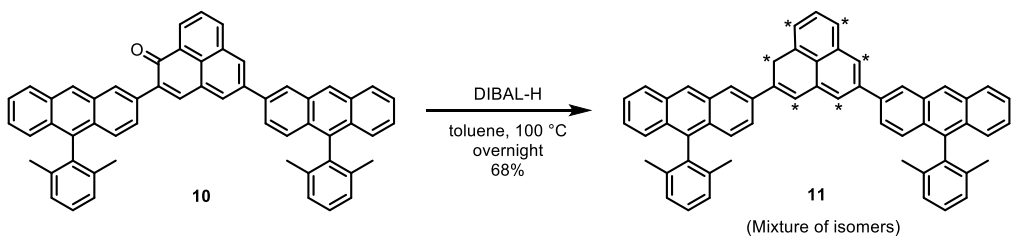

The compound was prepared using a modified literature procedure;[10] the reaction was carried out under inert conditions; solvents were deoxygenated by freeze-pump-thaw technique in three cycles under a nitrogen atmosphere. 2,5-Bis(10-(2,6-dimethylphenyl)anthracen-2-yl)-1*H*-phenalen-1-one (**10**; 9.3 mg, 0.020 mmol) was suspended in toluene (1 mL). A solution of DIBAL-H (25 µL, 0.030 mmol, 1.0 M in toluene) was added dropwise at room temperature and the reaction mixture was heated at 100 °C overnight. After cooling to room temperature, the reaction mixture was passed through a pad of Florisil® with toluene as eluent under the exclusion of air. After removal of the solvent in vacuum, the desired compound **11** was obtained as an orange solid (6.2 mg, 0.010 mmol, 68%). NMR: Compound **11** is obtained as a mixture of isomers where the methylene groups can be shifted to any *α*-position (asterisks) of the phenalenyl subunit. The mixture of isomers gives a complex proton NMR spectrum with characteristic singlets for the methylene groups. The ratio between methylene groups, aromatic protons, and methyl groups is 2:28:12 matching the expected structures. HRMS (EI) *m/z:* $[M]^+$ calcd. for $C_{57}H_{42}$: 726.3281, found: 726.3271.

**Synthesis of 2,2'-(10-(2,6-dimethylphenyl)anthracene-2,7-diyl)bis(1*H*-phenalen-1-one) (13)**

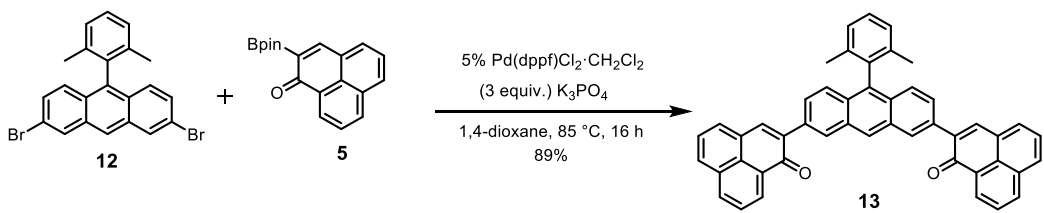

In a 25 mL Schlenk tube, compound **12** (9.5 mg, 22 µmol), compound **5** (20.0 mg, 65.3 µmol), Pd(dppf)Cl₂·CH₂Cl₂ (5.3 mg, 6.5 µmol), and K₃PO₄ (83 mg, 0.39 mmol) were added. The tube was evacuated and backfilled with argon three times. 1,4-Dioxane (5 mL) was then added, and the reaction mixture was stirred at 85 °C for 16 h under an argon atmosphere. After cooling to room temperature, the mixture was poured into water and extracted with dichloromethane (DCM) three times. The combined organic layers were washed with brine, dried over anhydrous MgSO₄, and concentrated under reduced pressure. The crude product was purified by silica gel column chromatography (iso-hexane/ethyl acetate = 3:1 v/v) to afford compound **13** as an orange solid (12 mg, 89% yield). HR-MALDI-TOF MS (matrix: DCTB) *m/z:* $[M]^+$ calcd. for $C_{48}H_{30}O_2$: 638.2240, found: 638.2258. ¹H NMR (300 MHz, 1,1,2,2-tetrachloroethane-*d₂*): δ 8.69 (dd, *J* = 7.4, 1.2 Hz, 2H), 8.61 (s, 1H), 8.41 (d, *J* = 1.7 Hz, 2H), 8.24 (dd, *J* = 8.2, 1.2 Hz, 2H), 8.08–7.99 (m, 4H), 7.90–7.78 (m, 4H), 7.67–7.57 (m, 4H), 7.49 (d, *J* = 9.0 Hz, 2H), 7.36

(dd, $J$ = 8.5, 6.3 Hz, 1H), 7.27 (d, $J$ = 7.5 Hz, 2H), 1.81 (s, 6H). $^{13}$C NMR (76 MHz, 1,1,2,2-tetrachloroethane-$d_2$): $\delta$ 184.3, 140.3, 138.9, 137.9, 137.3, 135.4, 135.2, 133.5, 132.0, 131.8, 131.7, 131.1, 129.8, 129.3, 129.0, 128.0, 127.8, 127.6, 127.5, 127.3, 127.2, 127.0, 125.7, 74.2, 20.3 (one carbon resonance could not be resolved because of signal overlap).

### Synthesis of 10-(2,6-dimethylphenyl)-2,7-di(1*H*-phenalen-2-yl)anthracene (14) (Mixture of isomers)

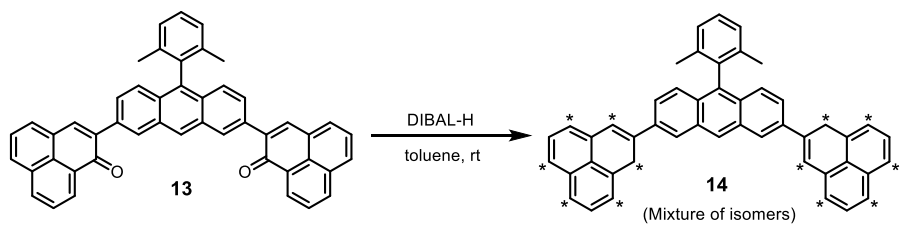

**Method A**: In a 10 mL Schlenk tube connected to a vacuum trap (Figure S32), compound **13** (5.0 mg, 7.8 µmol) was dissolved in dry, deoxygenated toluene (3 mL). Diisobutylaluminum hydride (DIBAL-H; 62.6 µL, 62.6 µmol, 1 M in hexane) was added dropwise at room temperature. Upon addition, the reaction mixture immediately changed color from orange to light yellow and exhibited sky-blue fluorescence under 365 nm UV light. The mixture was stirred at room temperature overnight. After completion, the reaction was carefully quenched with methanol and the solvent was removed *in situ* under reduced pressure. The residue was transferred into a glovebox, washed with dichloromethane (DCM), and filtered through a syringe filter. The DCM filtrate was then evaporated under reduced pressure to afford the crude isomeric mixture of compound **14**. HR-MALDI-TOF MS (matrix: dithranol) *m/z*: [*M*]$^+$ calcd. for C$_{48}$H$_{34}$: 610.2655, found: 610.2185. High-resolution MALDI-TOF mass spectrometry of the crude product showed an additional peak, likely resulting from partial oxidation of compound **14** during measurement (Figure S32). The resulting mixture was used directly as the precursor for the on-surface synthesis.

### Alternative synthetic procedure for compound 14 (Mixture of isomers)

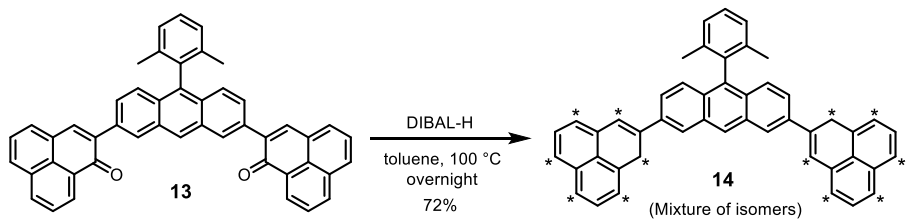

**Method B**: The compound was prepared using a modified literature procedure;[10] the reaction was carried out under inert conditions; solvents were deoxygenated by freeze-pump-thaw technique in three cycles under a nitrogen atmosphere. 2,2'-(10-(2,6-Dimethylphenyl)anthracene-2,7-diyl)bis(1*H*-phenalen-1-one) (**13**; 7.0 mg, 0.010 mmol) was suspended in toluene (1 mL). A solution of DIBAL-H (42 µL, 0.040 mmol, 1.0 M in toluene) was added

dropwise at room temperature and the reaction mixture was heated at 100 °C overnight. After cooling to room temperature, the reaction mixture was passed through a pad of Florisil® with toluene as an eluent under the exclusion of air. After removal of the solvent in vacuum, the desired compound **14** was obtained as yellow solid (4.6 mg, 0.010 mmol, 72%). NMR: Compound **14** is obtained as a mixture of isomers where the methylene groups can be shifted to any $\alpha$-position (asterisks) of the phenalenyl subunit. The mixture of isomers gives a complex proton NMR spectrum with characteristic singlets for the methylene groups. The ratio between methylene groups, aromatic protons, and methyl groups is 4:24:6 matching the expected structures. HRMS (EI) *m/z:* [*M*]$^+$ calcd. for $C_{48}H_{34}$: 610.2655, found: 610.2641.

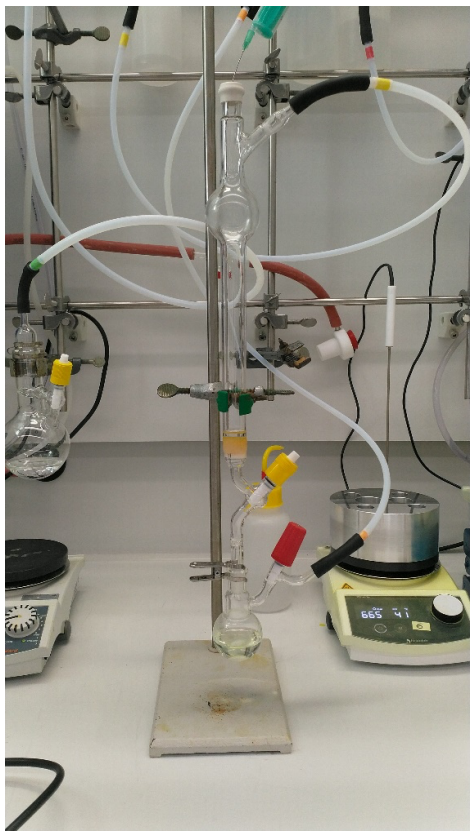

**Figure S9**. Custom-made glassware for inert column chromatography.

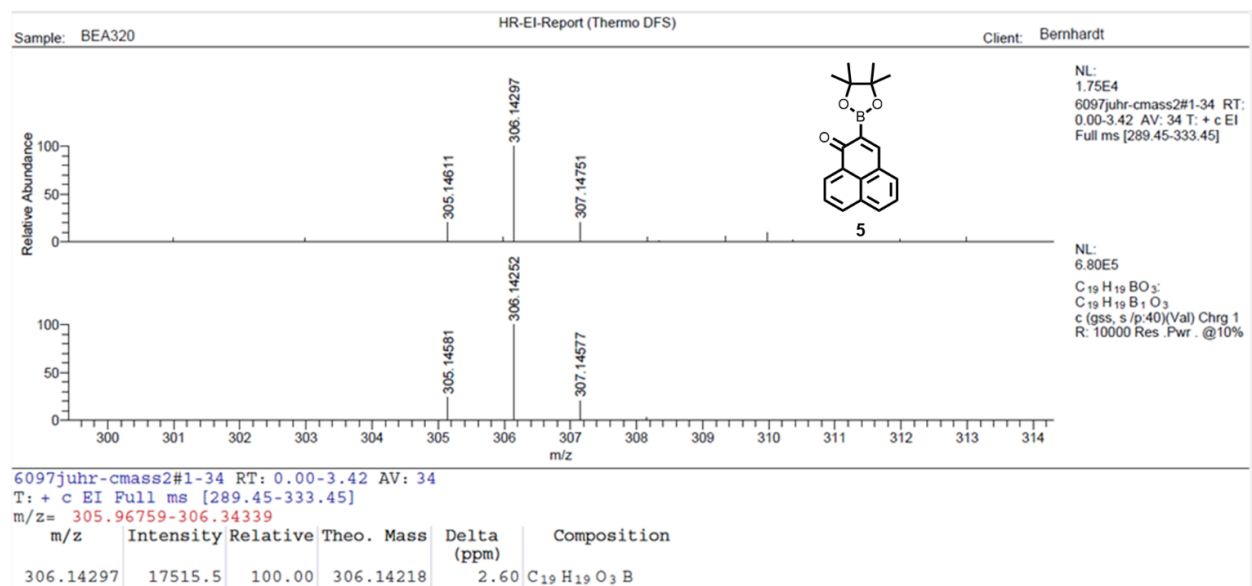

**Figure S10**. HR-EI-MS report of compound **5**.

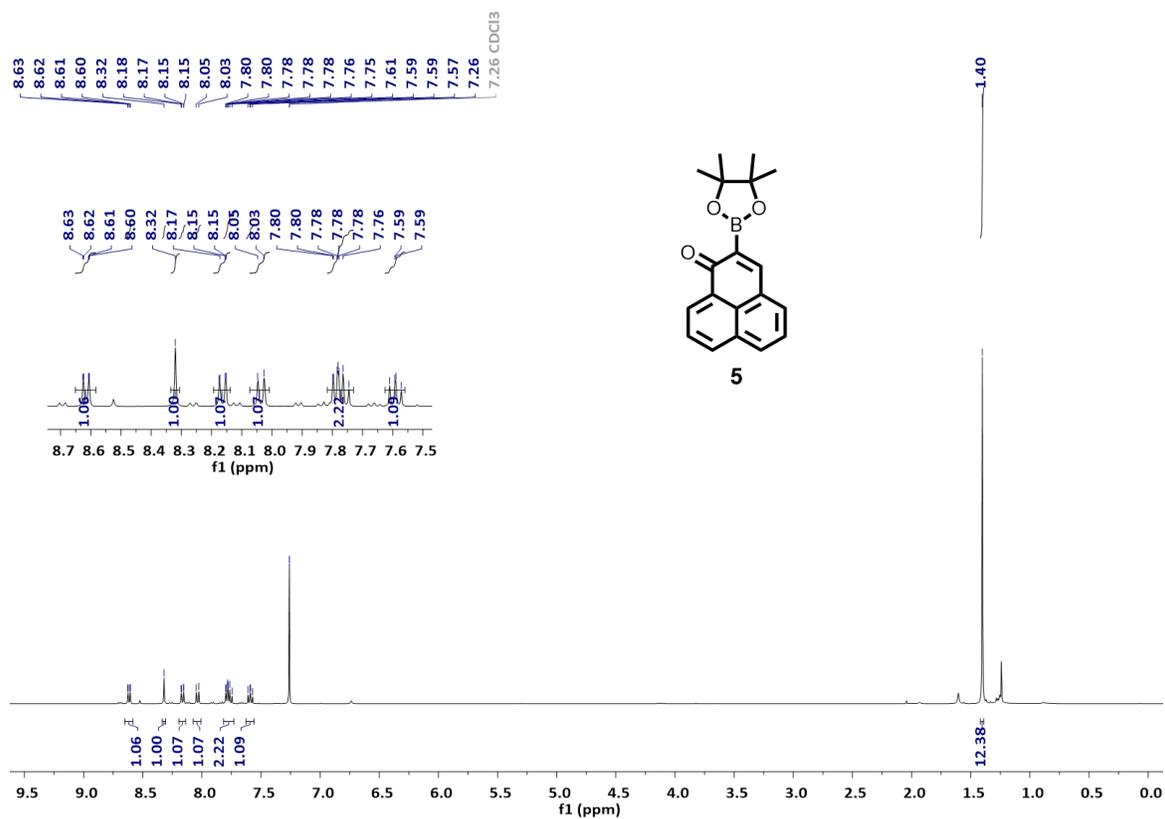

**Figure S11**. ¹H-NMR spectrum of **5** dissolved in CDCl₃, 400 MHz, 298 K.

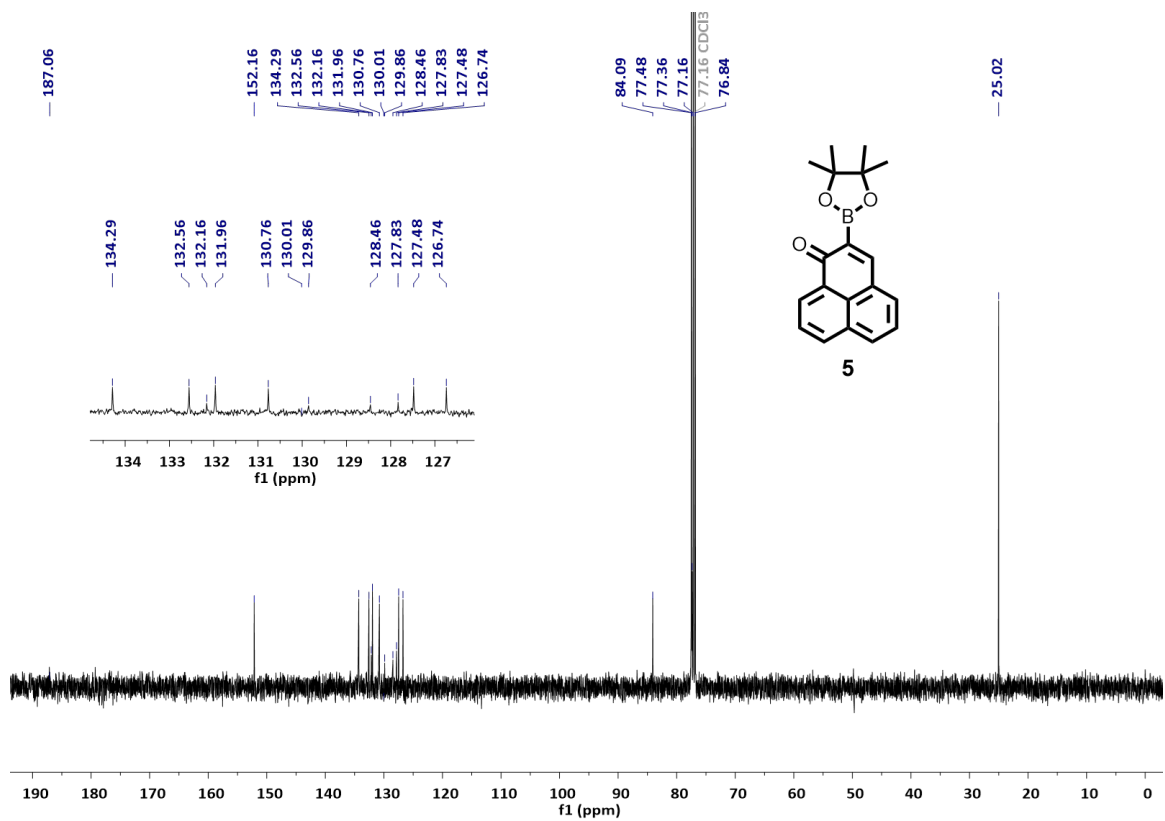

**Figure S12**. $^{13}$C-NMR spectrum of **5** dissolved in CDCl$_3$, 101 MHz, 298 K.

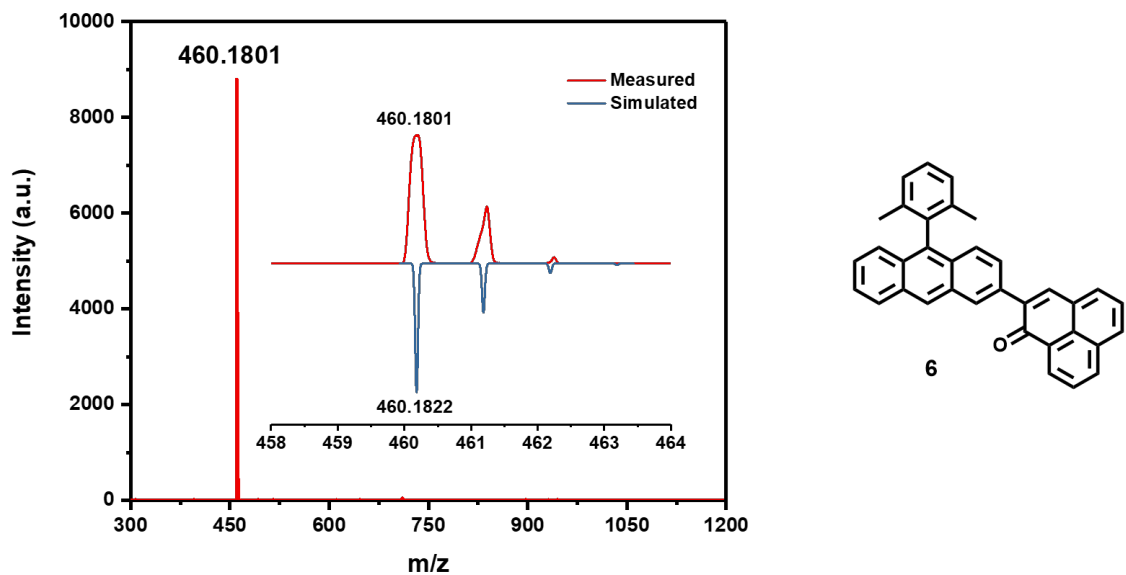

**Figure S13**. Liquid-state HR-MALDI-TOF-MS of **6** (matrix: DCTB).

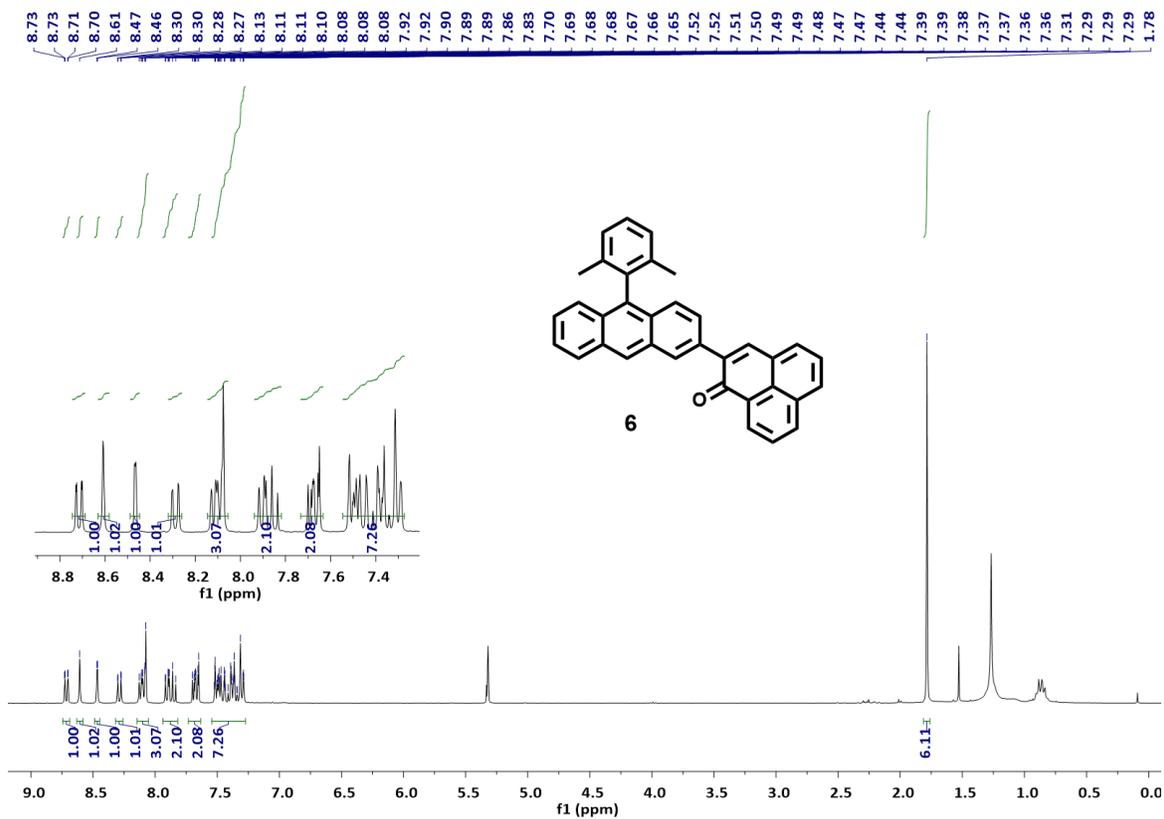

**Figure S14**. $^1$H-NMR spectrum of **6** dissolved in CD$_2$Cl$_2$, 300 MHz, 296 K.

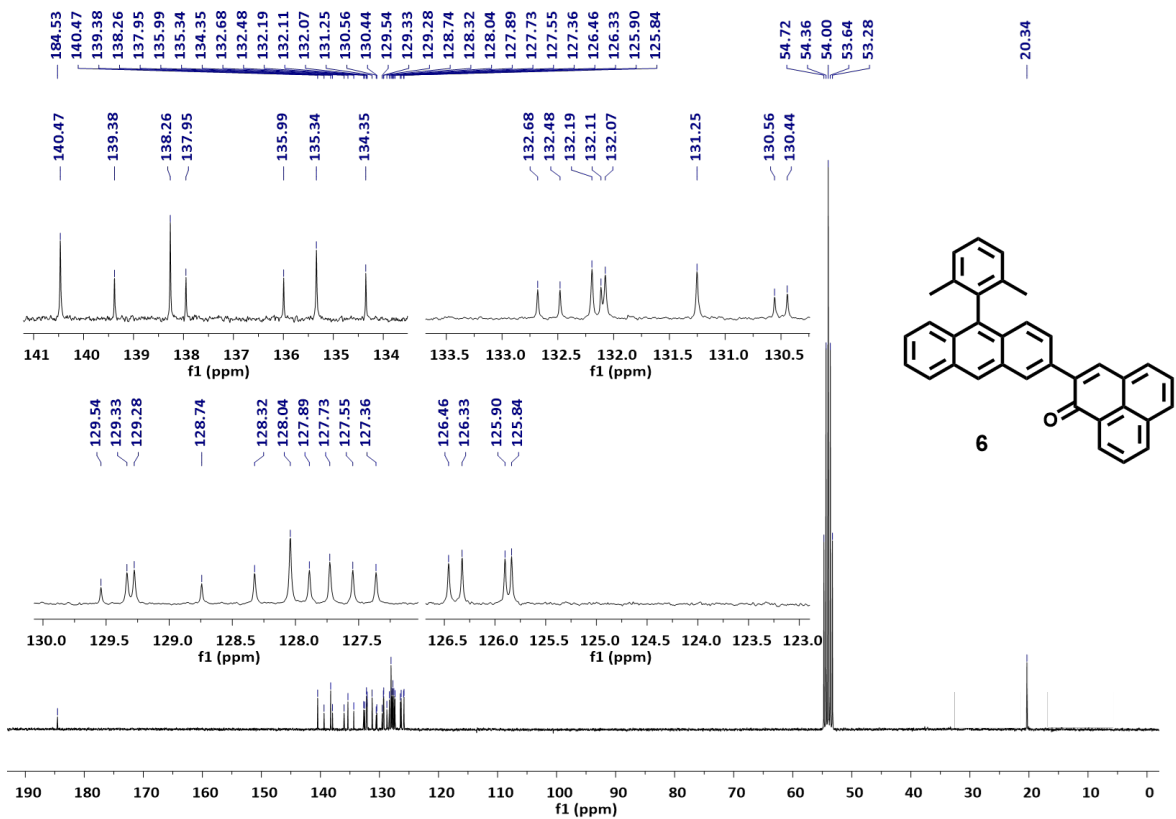

**Figure S15.** $^{13}$C-NMR spectrum of **6** dissolved in $CD_2Cl_2$, 76 MHz, 296 K (one carbon resonance could not be resolved because of signal overlap).

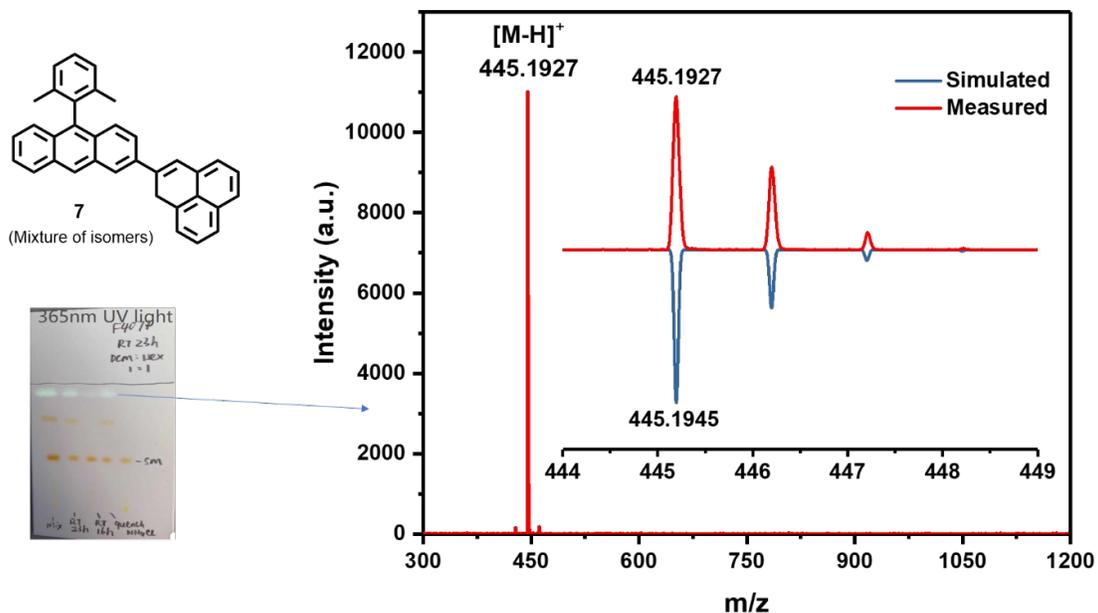

**Figure S16.** Compound **7** exhibited a distinct spot on TLC, showing sky-blue fluorescence under 365 nm UV light. And the liquid-state high-resolution MALDI-TOF mass spectrum of compound **7** (matrix: dithranol).

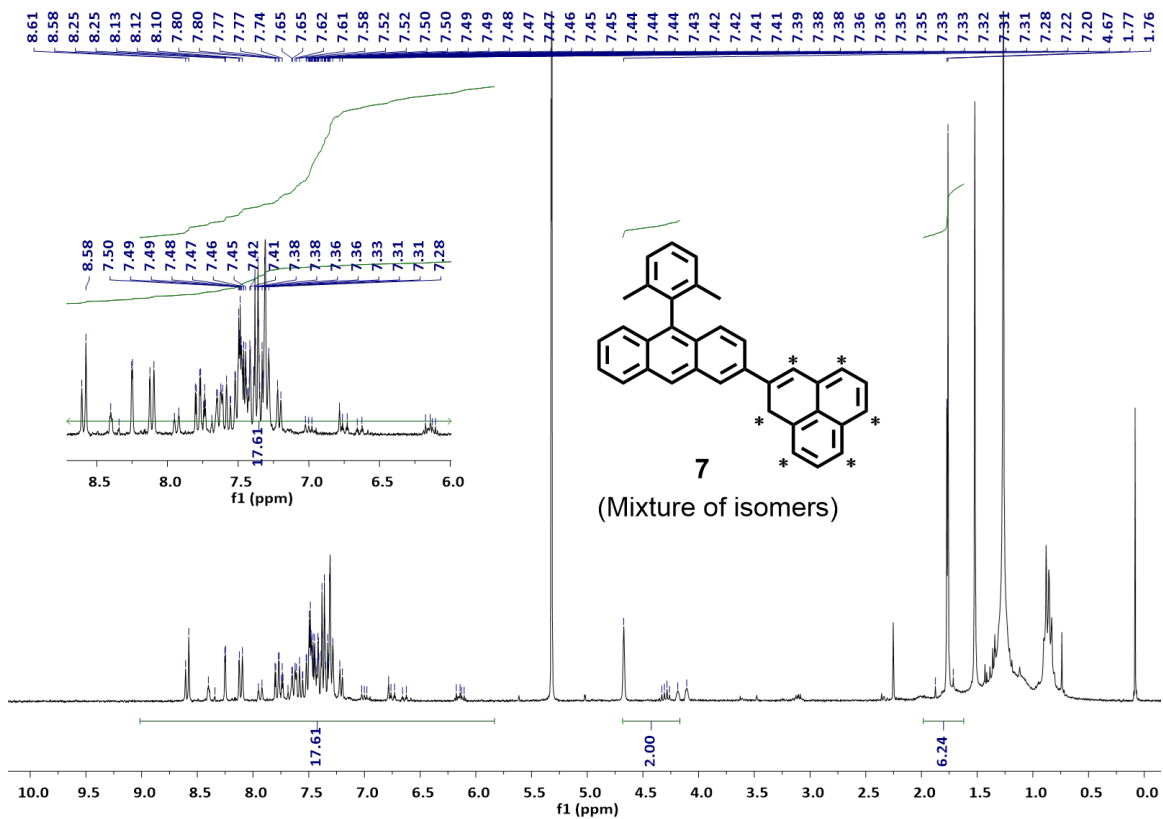

**Figure S17.** $^1$H-NMR spectrum of **7** dissolved in CD$_2$Cl$_2$, 300 MHz, 296 K.

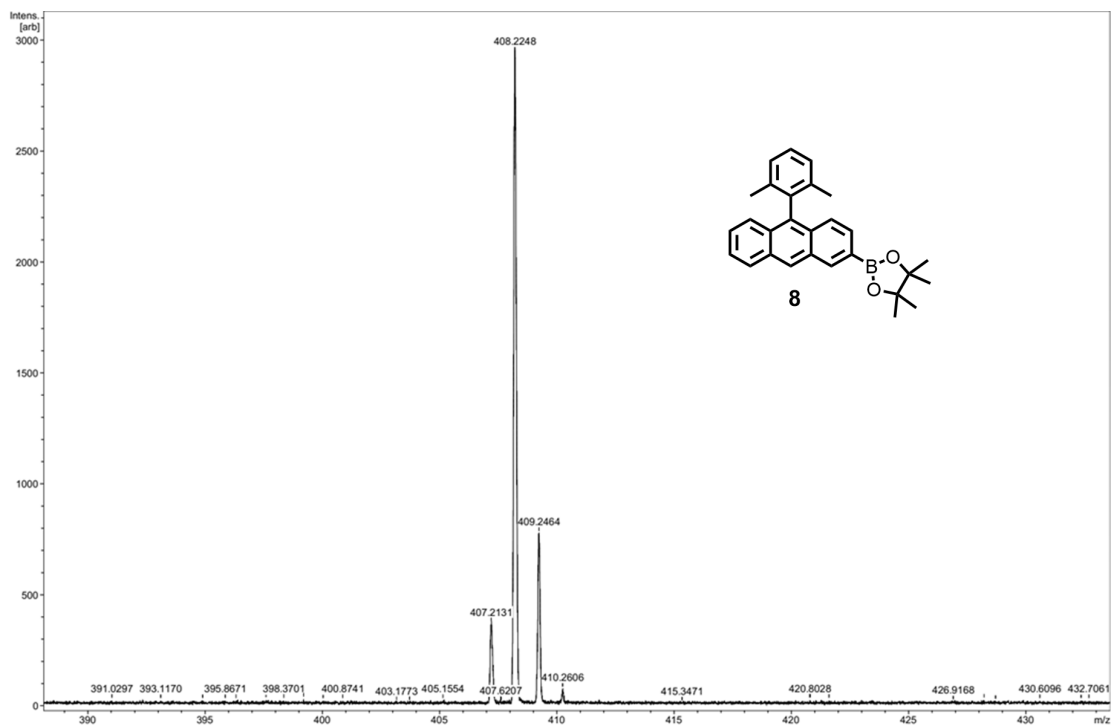

Bruker Daltonics flexControl

Display Screenshot - Generated On 2022-01-09 14h32m55s

**Figure S18**. Liquid-state HR-MALDI-TOF-MS of **8** (matrix: dithranol).

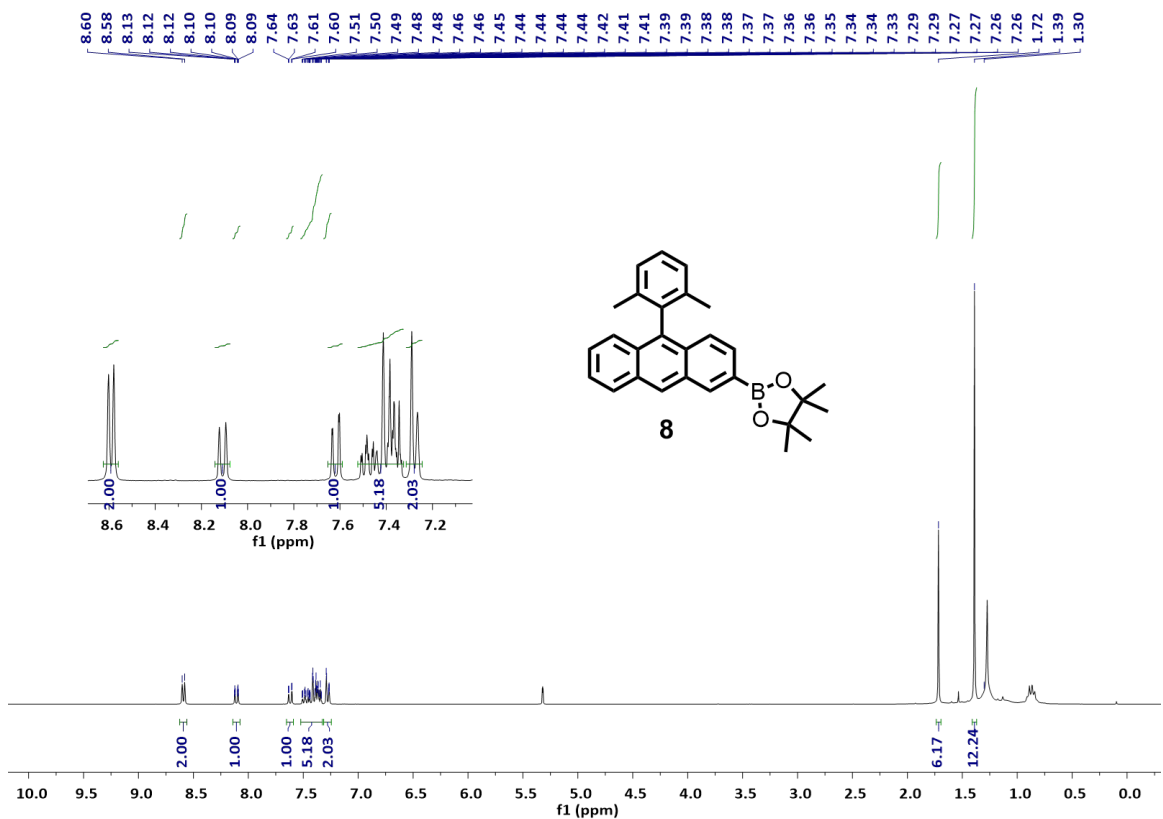

**Figure S19.** $^1$H-NMR spectrum of **8** dissolved in CD$_2$Cl$_2$, 300 MHz, 296 K.

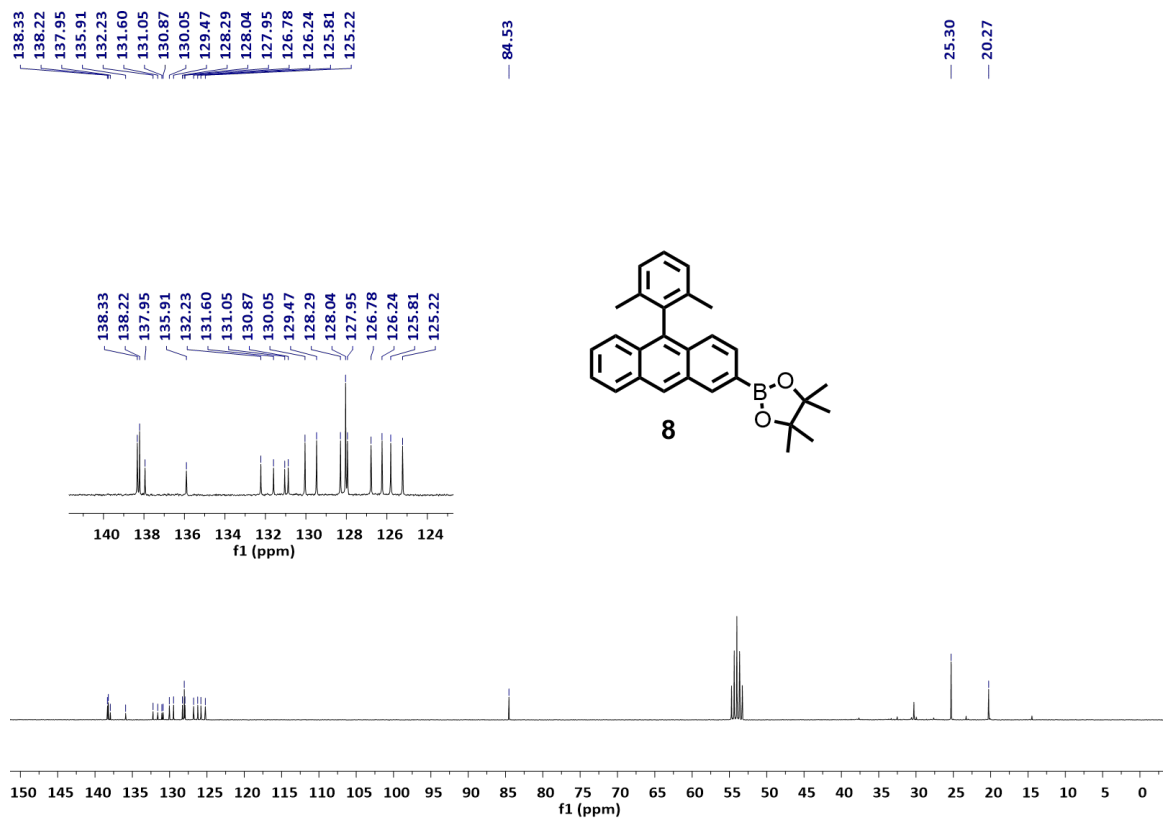

**Figure S20.** $^{13}$C-NMR spectrum of **8** dissolved in CD$_2$Cl$_2$, 76 MHz, 296 K (one carbon resonance (presumably C–B) could not be detected).

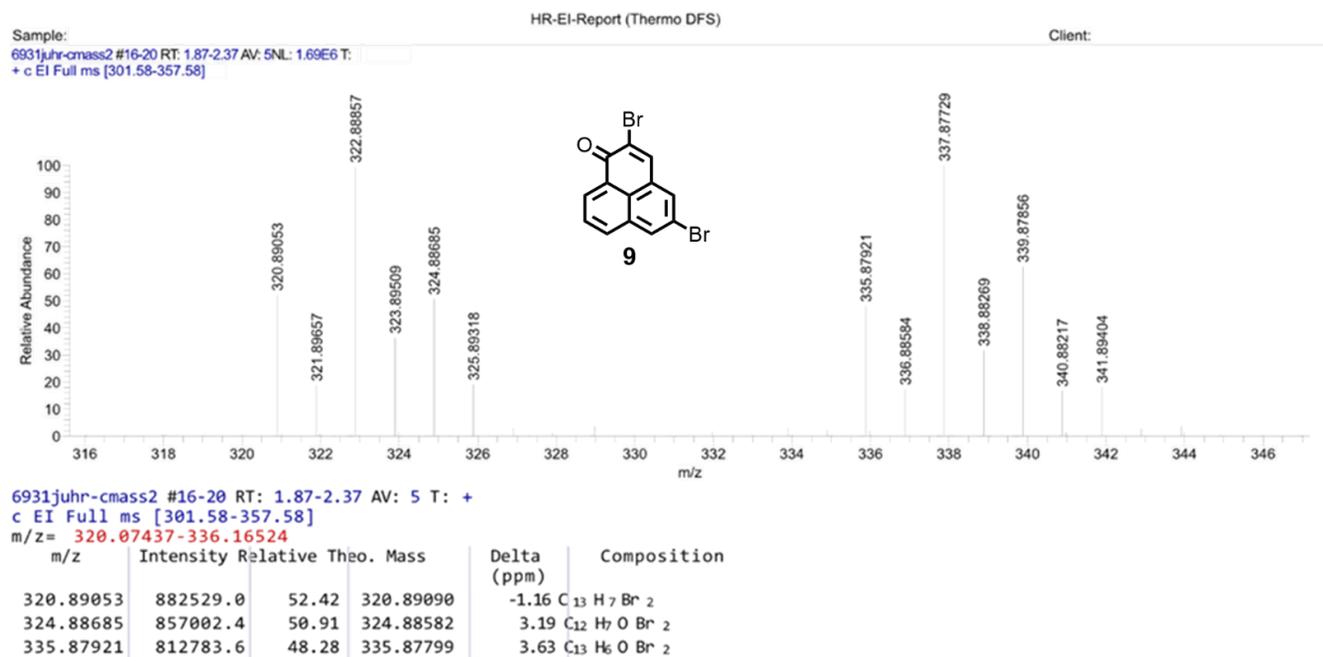

**Figure S21.** HR-EI-MS report of compound **9**.

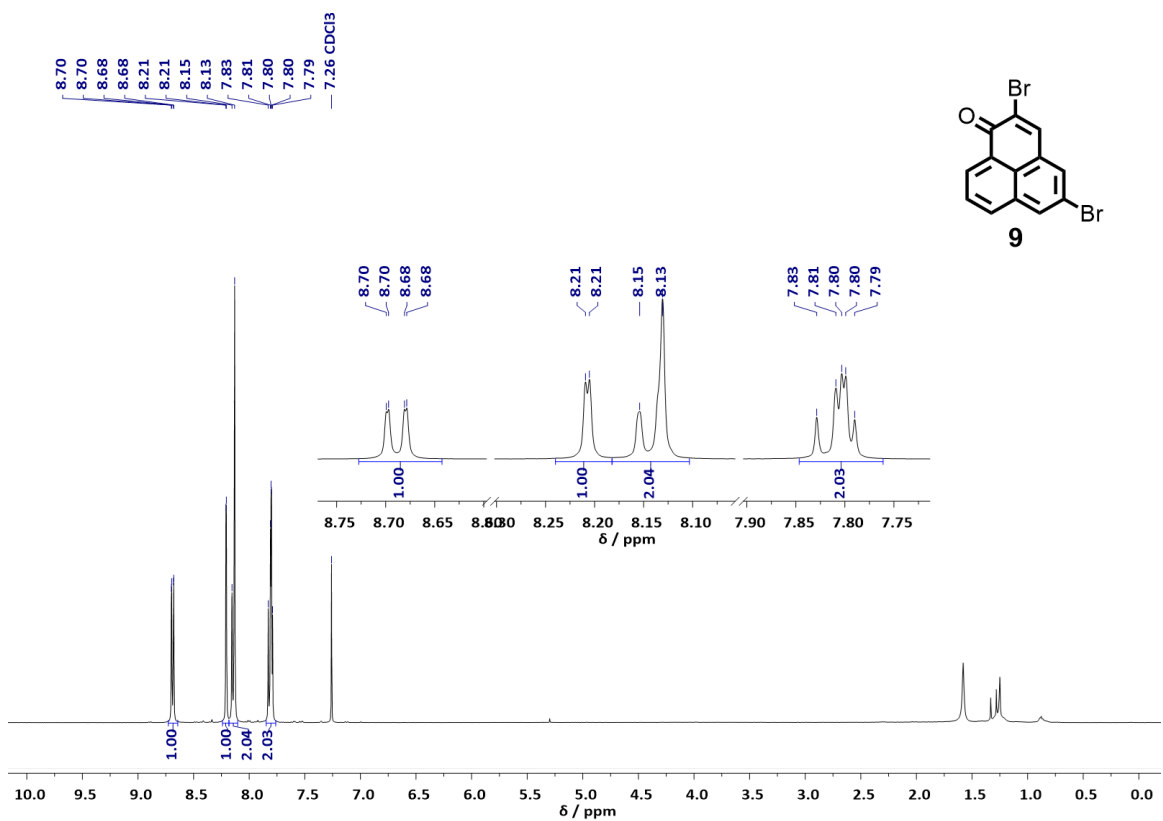

**Figure S22**. $^1$H-NMR spectrum of **9** dissolved in CDCl$_3$, 400 MHz, 298 K.

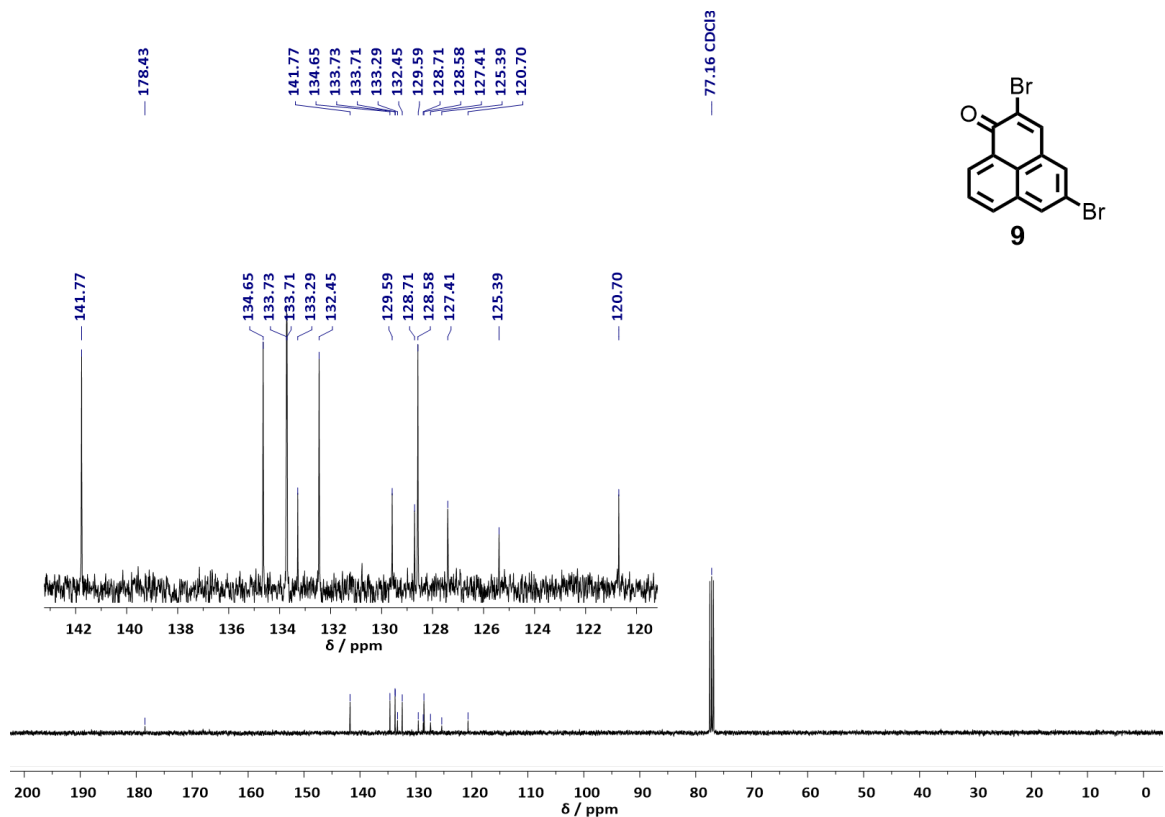

**Figure S23**. $^{13}$C-NMR spectrum of **9** dissolved in CDCl$_3$, 101 MHz, 298 K.

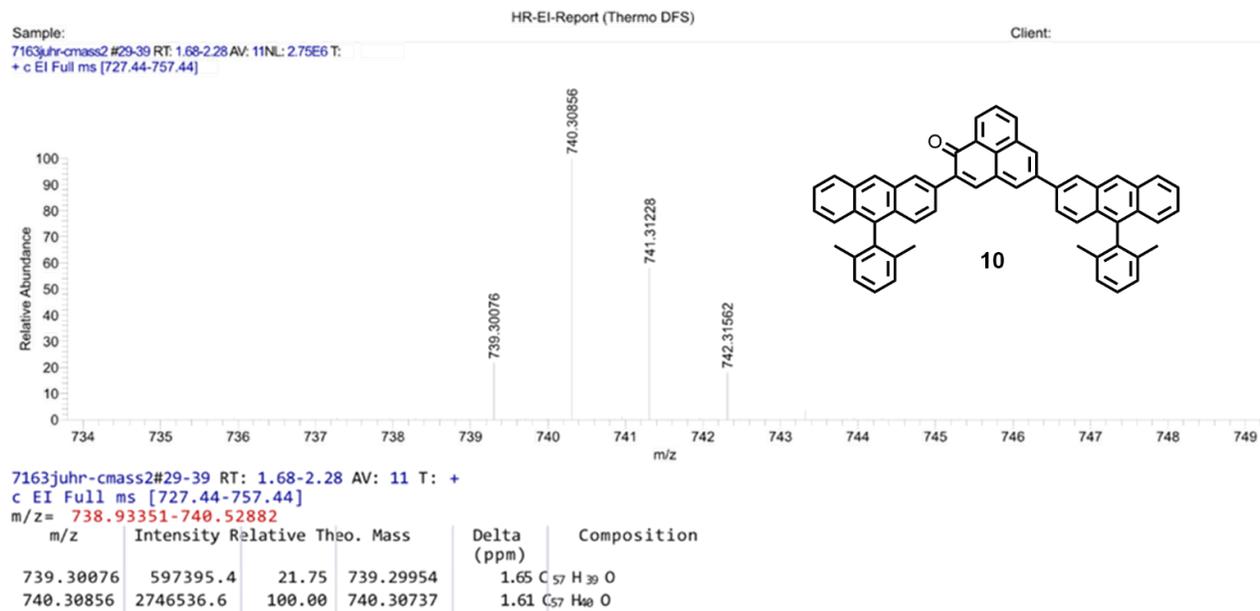

**Figure S24**. HR-EI-MS report of compound **10**.

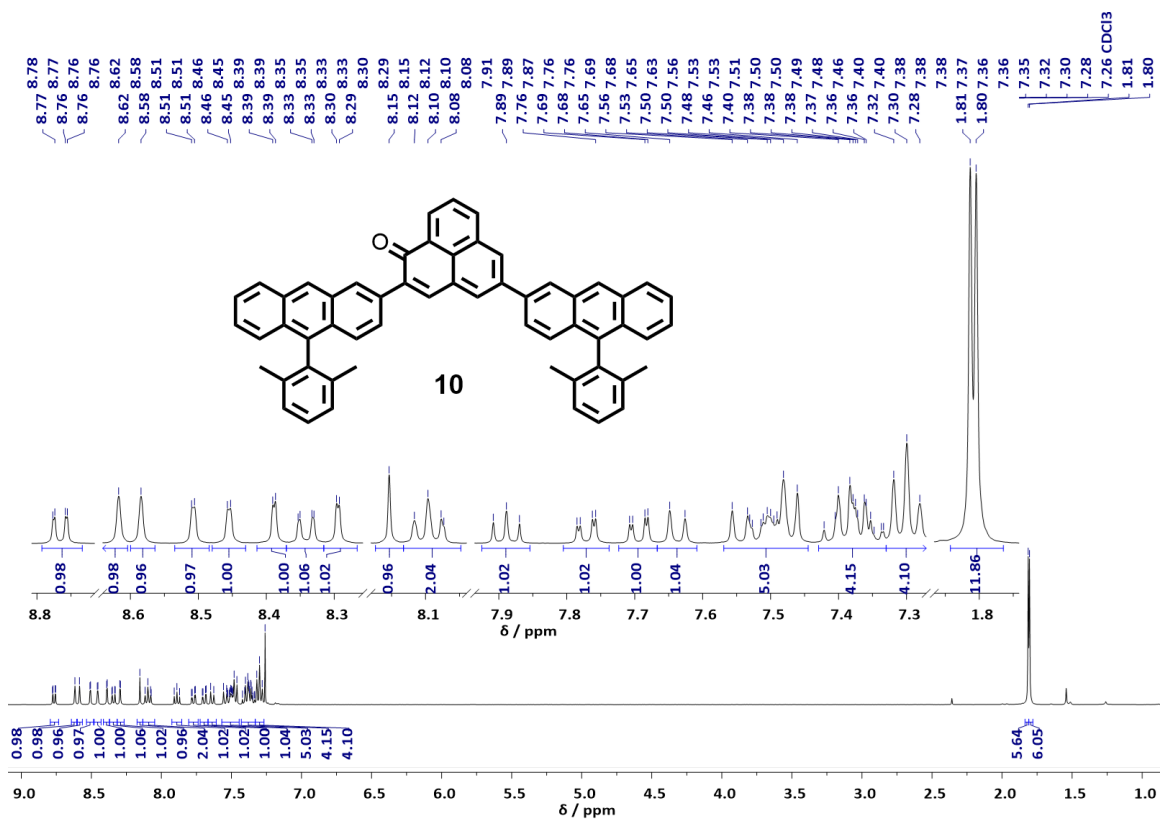

**Figure S25**. ¹H-NMR spectrum of **10** dissolved in CDCl₃, 400 MHz, 298 K.

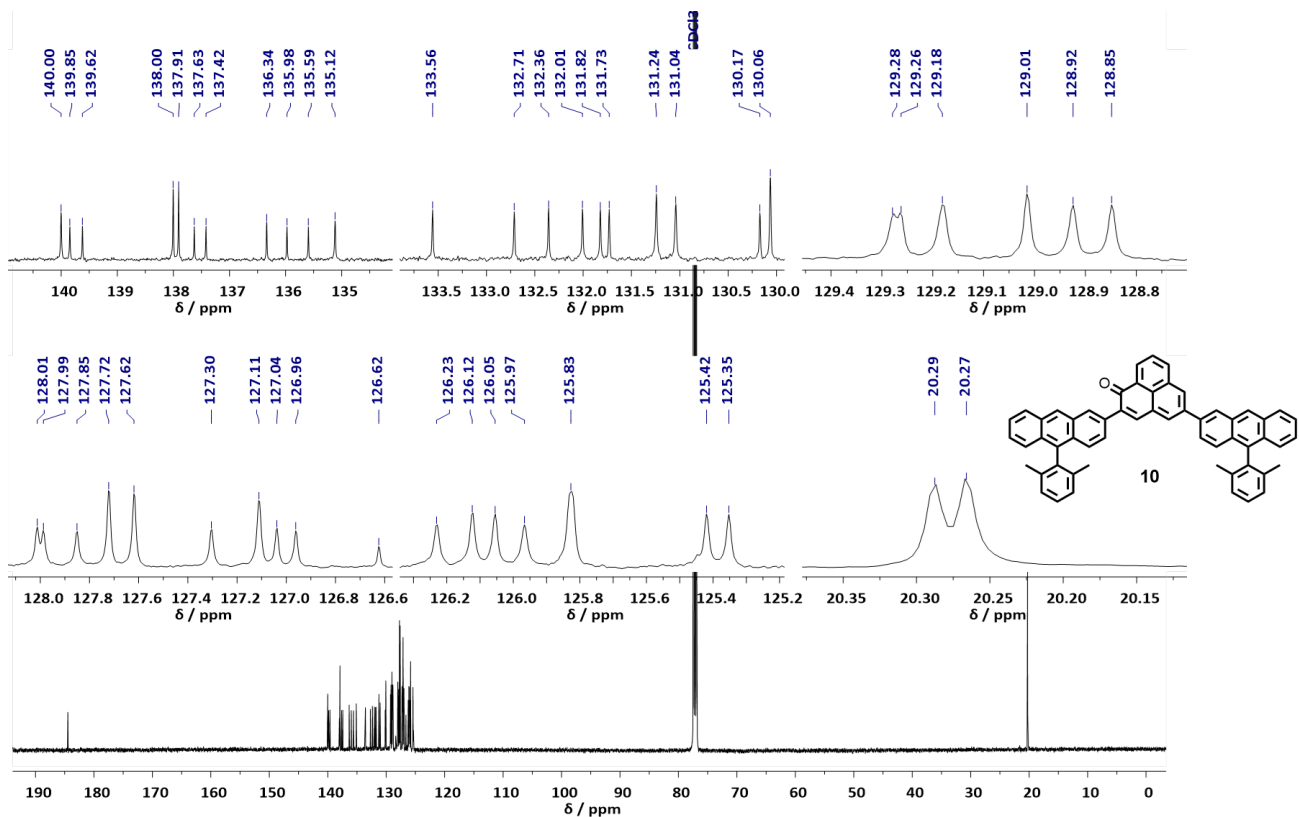

**Figure S26**. ¹³C-NMR spectrum of **10** dissolved in CDCl₃, 101 MHz, 298 K (four carbon resonances could not be resolved because of signal overlap).

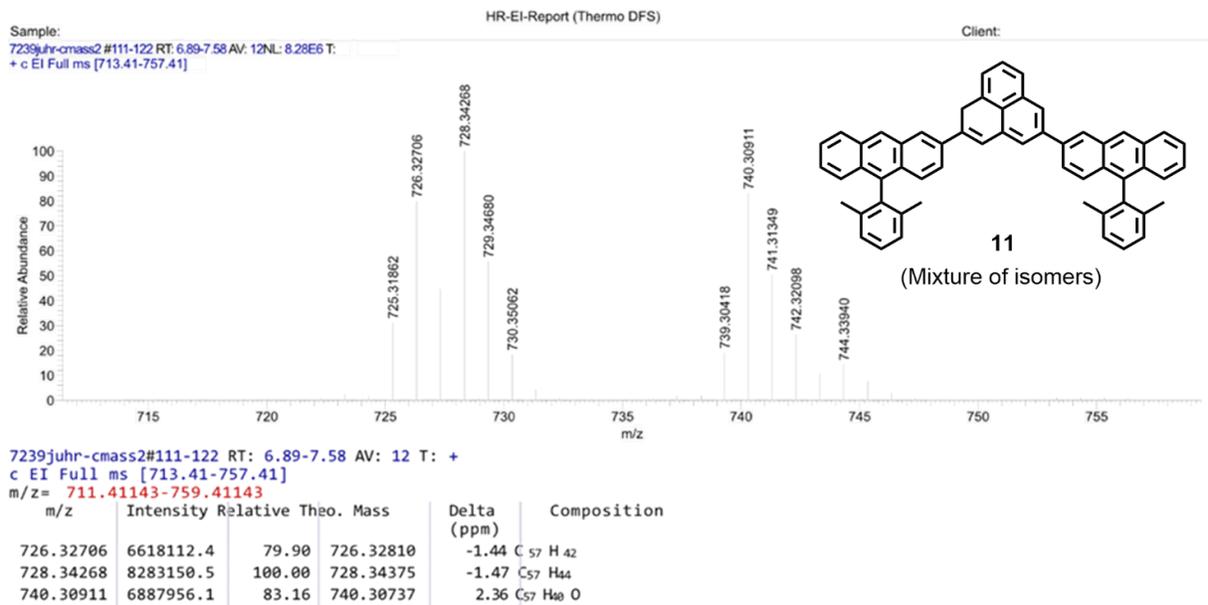

**Figure S27**. HR-EI-MS report of compound **11**.

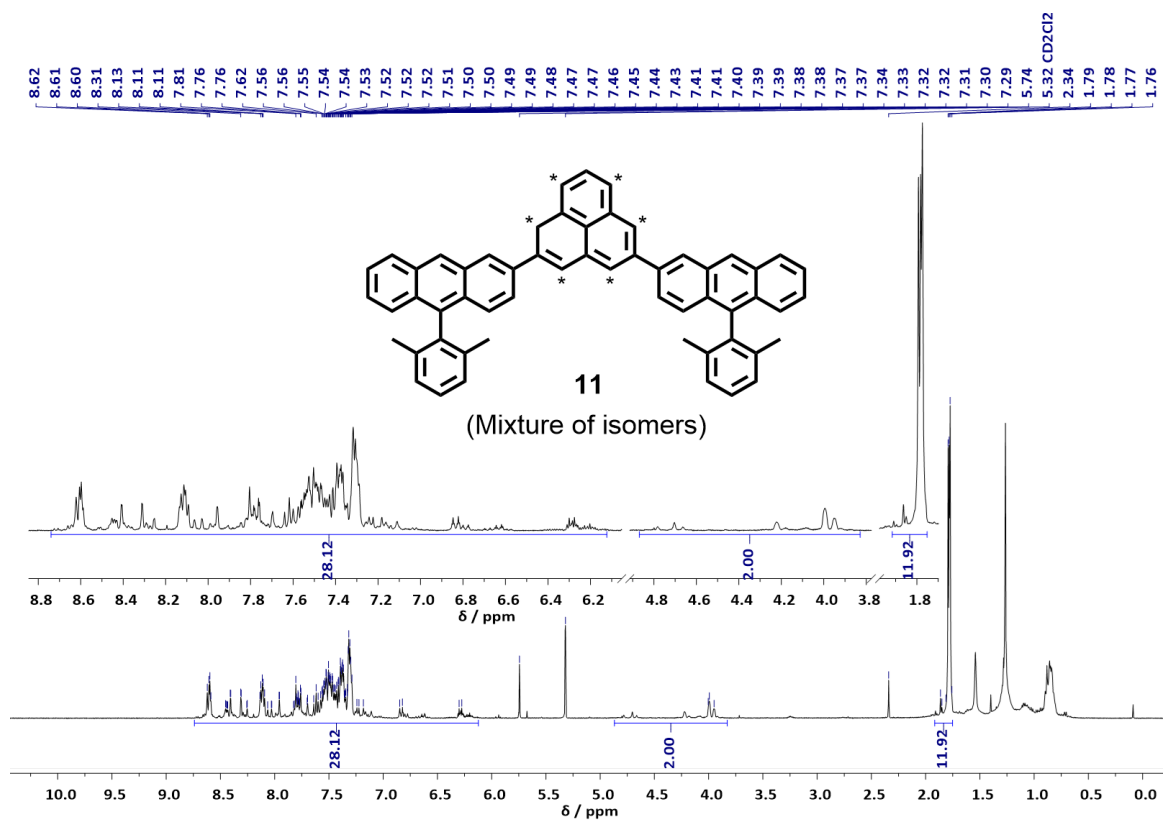

**Figure S28**. ¹H-NMR spectrum of **11** dissolved in CD₂Cl₂, 400 MHz, 298 K.

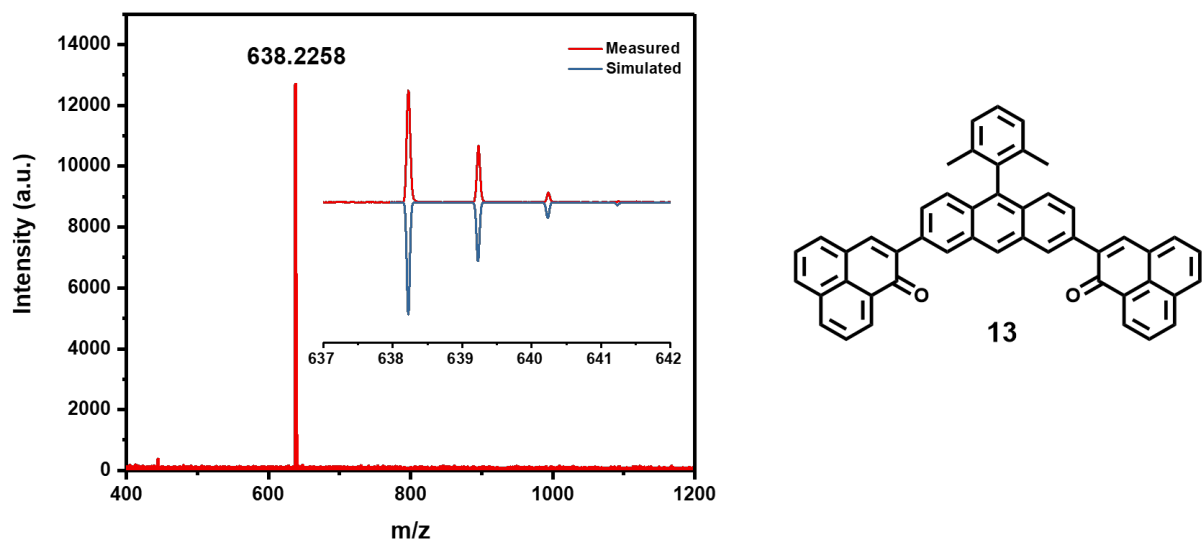

**Figure S29**. Liquid-state HR-MALDI-TOF-MS spectrum of compound **13** (matrix: DCTB).

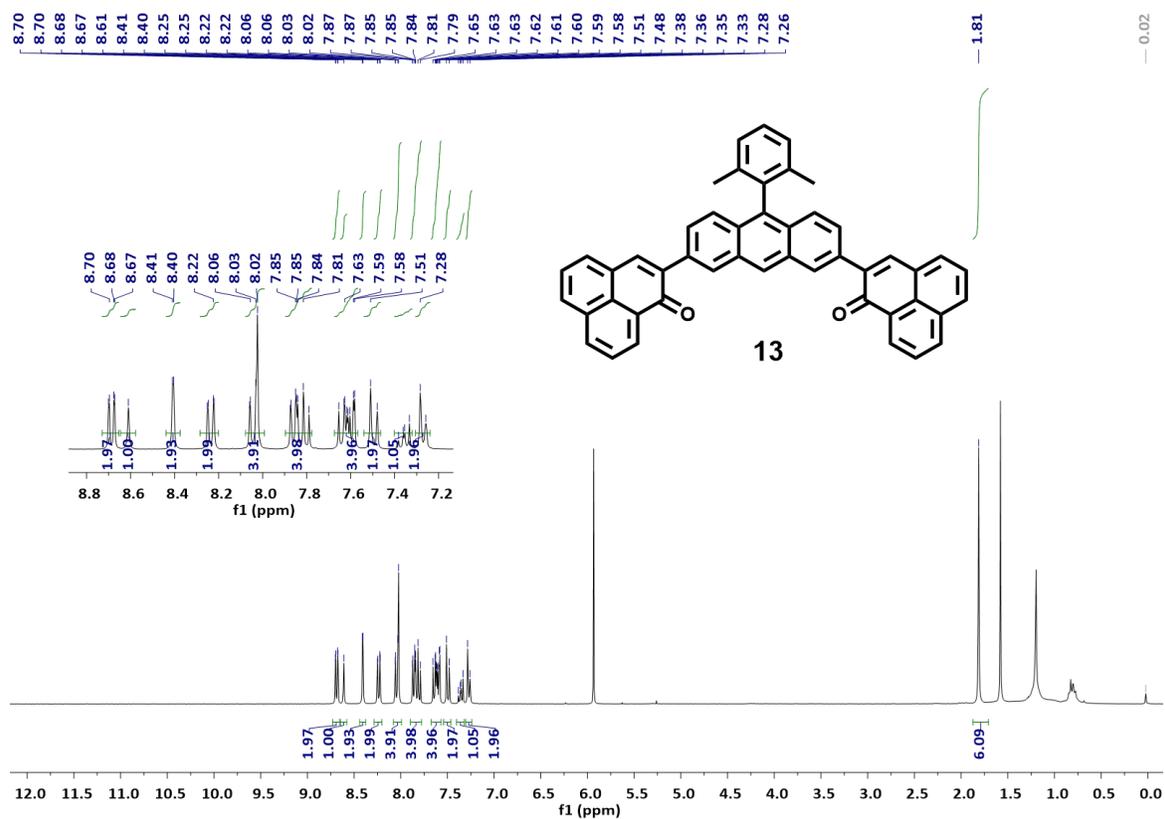

**Figure S30**. $^{1}$H-NMR spectrum of compound **13** dissolved in 1,1,2,2-tetrachloroethane-$d_2$, 300 MHz, 296 K.

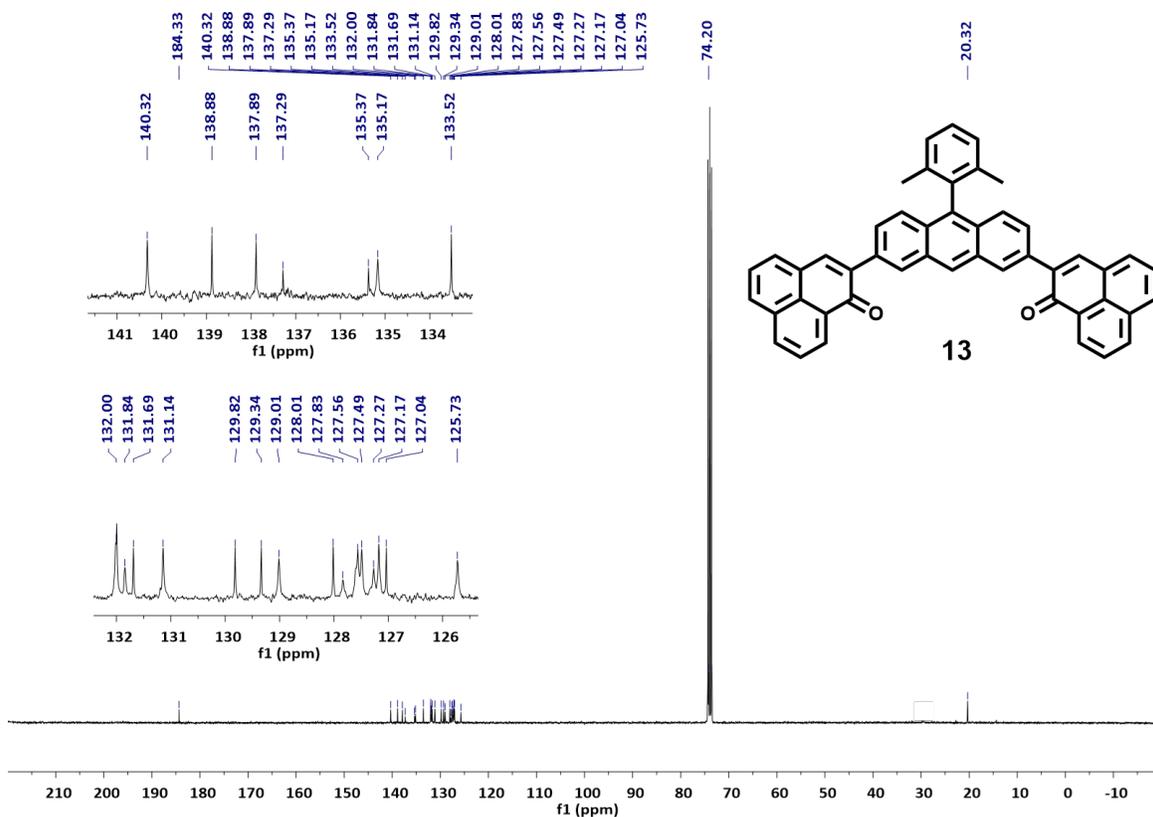

**Figure S31**. $^{13}$C-NMR spectrum of **13** dissolved in 1,1,2,2-tetrachloroethane-$d_2$, 75 MHz, 296 K (one carbon resonance could not be resolved because of signal overlap).

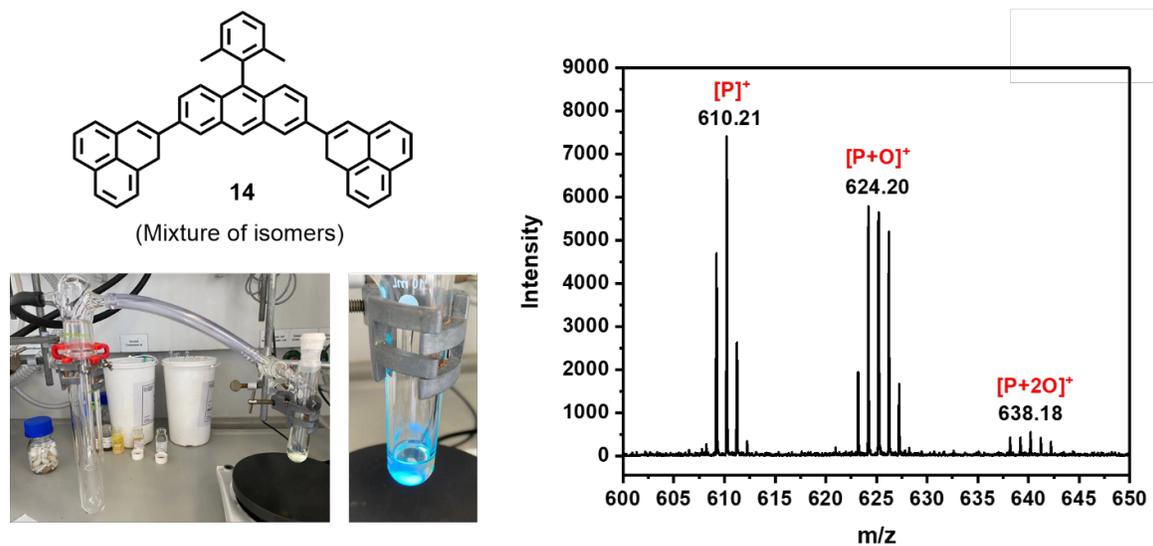

**Figure S32**. Method A: Experimental setup for the reduction reaction leading to compound **14**, along with the MALDI-TOF mass spectrum of the crude product (matrix: dithranol)..

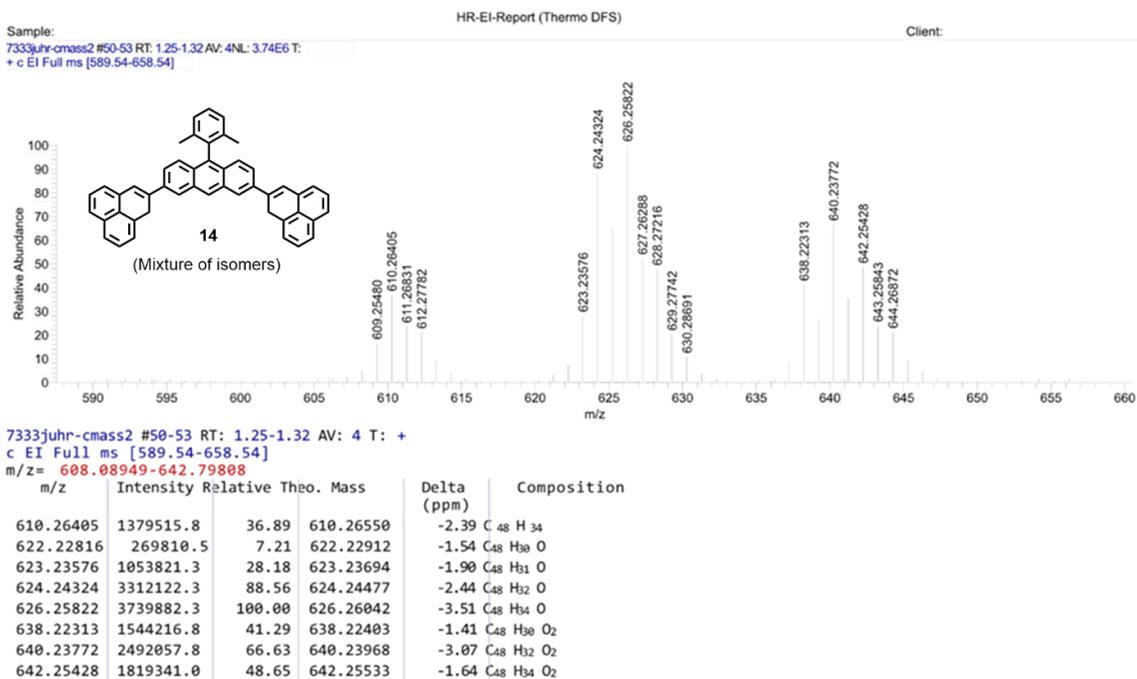

**Figure S33**. Method B: HR-EI-MS report of compound **14**.

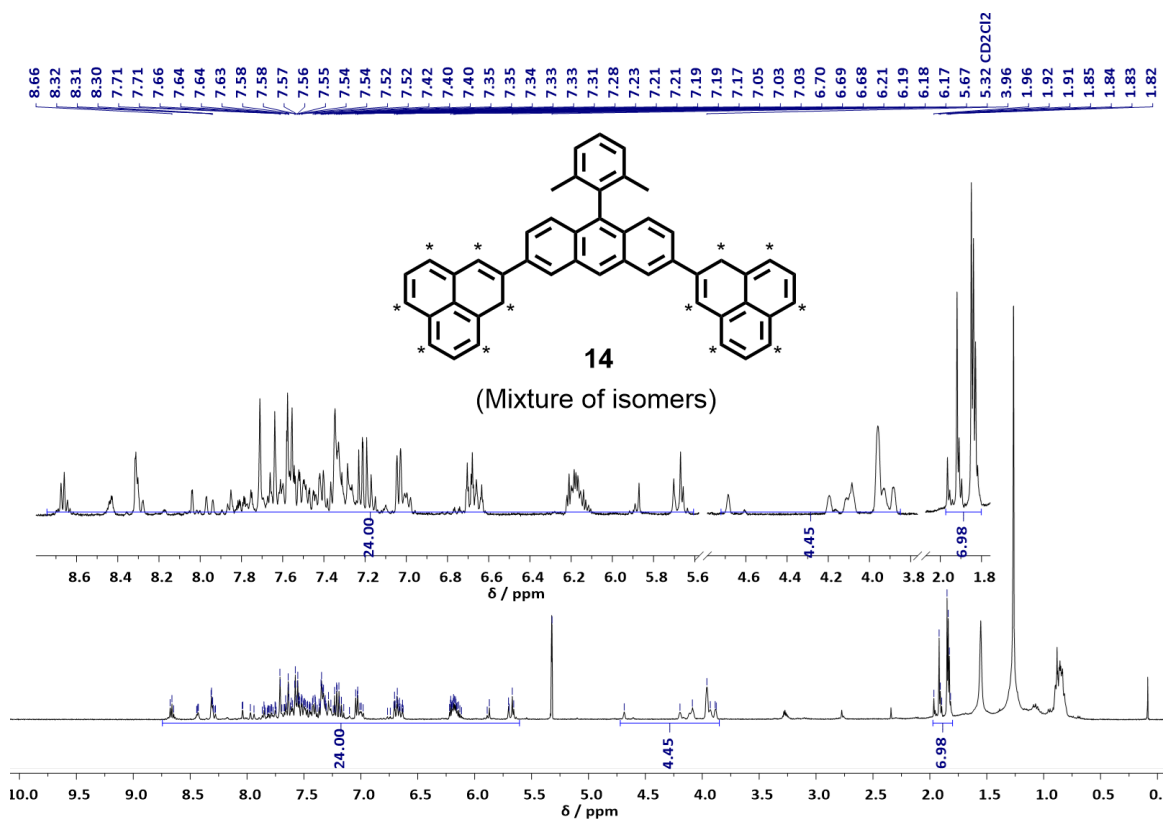

**Figure S34**. $^1$H-NMR spectrum of **14** dissolved in CD$_2$Cl$_2$, 400 MHz, 298 K.



%\bibliography{references,footnotes} % Produces the bibliography via BibTeX.